%% file: main.tex
\documentclass[article]{IEEEtran}
\input{prologue}
\begin{document}
	\title{GSP = DSP + Boundary Conditions\\ The Graph Signal Processing Companion Model}
\author{John Shi, \IEEEmembership{Member, IEEE,} Jos\'e M.~F.~Moura, \IEEEmembership{Fellow, IEEE}%
\thanks{This work is partially supported by NSF grant CCF-2327905.}%
\thanks{Department of Electrical and Computer Engineering, Carnegie Mellon University, Pittsburgh PA 15217 USA; [jshi3,moura]@andrew.cmu.edu.}}

\maketitle
\begin{abstract}
This paper considers a graph signal processing (GSP) model, the GSP companion model, that explores the cyclic nature of the GSP signal space, i.e., that, under appropriate conditions, there is a cyclic generator for the signal space. The paper applies to directed or undirected graphs, whether we work with the adjacency matrix or the graph Laplacian. This reflects the cyclic nature of the time digital signal processing model. The companion model replicates the DSP model, except for different assumed boundary conditions (bc), metaphorically, GSP=DSP+bc.  Based on this, we propose the Step~1-Step~2 Companion Model Method (S$^2$CM$^2$) to expand new concepts to GSP, namely, concepts that have proven useful in DSP. The paper illustrates the application of S$^2$CM$^2$ by showing how to define the GSP $z$-transform, that GSP convolution in the companion model can be achieved with the FFT,  and by showing that GSP modulating carriers are not the graph eigenvectors but appropriate (Hadamard) powers of the graph eigenfrequency vector that collects all graph frequencies. In an appendix, the papers considers the numerical computation of the companion model using Lagrange barycentric interpolation.
\end{abstract}
\vspace{-.3cm}
\section{Introduction}\label{sec:intro}
Graph signal processing~(GSP) \cite{Sandryhaila:13,Sandryhaila:14,ShumanNFOV:13} extends digital signal processing~(DSP) \cite{oppenheimwillsky-1983,siebert-1986,oppenheimschaffer-1989,mitra-1998} from time signals and images to data indexed by nodes of a graph $G=(V,E)$. New GSP concepts are usually introduced directly in what we will refer to in this paper as the vertex or the spectral (graph frequency) domains. These are natural domains and reflect the corresponding DSP time and frequency dual domains. This approach has carried the field so far. This paper presents an alternative approach of extending GSP with new concepts and algorithms\textemdash the Step~1-Step~2 Companion Model Method (S$^2$CM$^2$). We advocate that, in Step~1, the novel concept or methodology be introduced in the \textit{companion} GSP model, see section~\ref{sec:gspcompanionmodel},  and then, using a transformation between the vertex or spectral domains and the companion domain, obtain the concept or new processing method in the traditional domains. The paper describes the companion GSP model and then illustrates application of S$^2$CM$^2$ with concepts like graph impulse, GSP $z$-transform, fast linear and circular convolution in the companion GSP model, and modulation. The companion model is closest to the DSP time model, so that bringing a new  concept to GSP is made easier by using directly the corresponding DSP concept. This however should be taken with a caveat\textemdash because graphs are finite objects, there are issues with boundary effects. Surprisingly, the companion GSP model shows how to address these boundary effects, revealing what are the boundary conditions (bc) that each particular GSP model assumes. Figuratively, we can then state that, in the companion model, GSP=DSP+bc. Besides providing a systematic way of extending DSP to GSP, the companion model also helps in making clear choices when there are plausible multiple possible versions of a GSP concept. Of course, we do not claim that S$^2$CM$^2$ will work at every instance and  a lot of effort and intuition will be needed to continue expanding GSP. But we do think that S$^2$CM$^2$ provides a methodology to quickly capture to GSP successful proven DSP methods.

The companion model adds to our understanding of GSP in another fundamental way. The vertex GSP model uses the standard representation of a signal in terms of the standard, Euclidean vectors $\mathbf{e}_n$. This is natural, since each $\mathbf{e}_n$ gives the data in vertex~$n$. The spectral representation decomposes the signal space into irreducible invariant spaces\footnote{As explained later, irreducible means that the space cannot be further decomposed in invariant subspaces. Invariant is with respect to linear maps that in GSP are the filters.}. The companion GSP model, as we show, considers the cyclic nature of the signal space. Under appropriate assumptions, the signal space is generated by a single generator, i.e., by a vector and its shifts by the filter under study.

\textbf{Brief review of the literature}. The GSP literature is vast, covering many topics in processing graph signals. In \cite{Sandryhaila:13,Sandryhaila:14,Sandryhaila:14big}, building on the Algebraic Signal Processing  \cite{Pueschel:03a,Pueschel:05e,Pueschel:08a,Pueschel:08b,Pueschel:08c}, the shift~$A$ is the basic  constituent block. It applies to generic directed or undirected graphs. The approach in \cite{ShumanNFOV:13} departs from a variational operator, the graph Laplacian~$L$, motivated by for example earlier work from spectral graph theory \cite{Chung:96,Belkin:02,Coifman:05a}, from work extending wavelets to data from irregularly placed sensors in sensor networks \cite{guestrinbodikthibauxpaskinmadden-ipsn2004,wagnerbaraniuketal-SSPWorkshop2005,Hammond:11,Narang:12}, and from research on sampling graph based data \cite{Narang:10,Narang:11}. It applies only to undirected graphs. A comprehensive review covering both approaches and illustrating many different applications of GSP is \cite{ortegafrossardkovacevicmouravandergheynst-2018}, see also the recent historical review of the field \cite{leusmarquesmouraortegashuman}, and the numerous references therein.

Many additional topics have been considered in GSP. A sample of these include: alternative (unitary, but not local) shift operators \cite{giraultgoncalvesfleury-2015,gavilizhang-2017}; approximating graph signals \cite{Zhu:12}; extensive work on sampling of graph signals, e.g., \cite{anis2014towards,Jelena,marques2015sampling,anis2016efficient,chamon2017greedy,tanaka-2018,shimoura-2021}, see the recent review \cite{EldarTanakaSPM}; extending classical multirate signal processing to graphs \cite{tekevaidyanathan-2017-I,tekevaidyanathan-2017-II}; an uncertainty principle for graph signals \cite{Agaskar:12}; the study of graph diffusions \cite{pasdeloupgriponmercierpastorrabbat-2018}; graph signal recovery \cite{chen2015signal-2}; interpolation and reconstruction of graph signals \cite{segarra2015interpolation,segarra2016reconstruction}; GSP modulation \cite{convolutionother}, stationarity of graph processes \cite{marquessegarraleusribeiro-17}; learning graphs from data \cite{meimoura-2017,meimoura-2018,dongthanourabbatfrossard-2019}; or non-diagonalizable shifts and the graph Fourier transform \cite{derimoura-2017}. These references describe graph signals by their vertex representation~$s$ or graph Fourier representation~$\widehat{s}$. None has discussed the companion model we present, nor the strategy and GSP extensions proposed here.

\textbf{Summary of the paper}. Section~\ref{sec:backgroundongsp} provides background on GSP, starting with the main tenets in algebraic signal model \cite{Pueschel:05e,Pueschel:08a,Pueschel:08b,Pueschel:08c}. Section~\ref{sec:vertexspectralmodels} reviews two GSP models, the vertex or standard model and the graph spectral model. For each model, we identify the algebra of filters, the signal module, and the choice of basis for the signal space, the shift, the graph Fourier transform, and the underlying graph.  Section~\ref{sec:gspcompanionmodel} introduces the companion model, by choosing a graph signal as the graph impulse. The section adopts this graph impulse and its shifts as the basis of the signal space, introducing the \textit{companion} basis, the companion signal representation, the companion shift, and the companion graph. The section studies the structure of the companion shift and how it reflects the boundary conditions assumed by the GSP model. Numerical examples compute the representations of a signal in the different domains. Section~\ref{sec:vertexspcompanion} contrasts the three GSP models, the vertex, spectral, and companion models.  Section~\ref{sec:compmodS2CM2} presents S$^2$CM$^2$ and illustrates its application. This section, besides considering the graph impulse, introduces the GSP $z$-transform, shows that graph convolution in the companion model is a fast $O(N\log N)$, and introduces a method for amplitude modulation and frequency division multiplexing of graph signals. This is illustrated with a simple example.  Section~\ref{sec:conclusion} concludes the paper. The companion model and the results of this paper apply to directed or undirected graphs, with some  examples using directed graphs while others undirected graphs, and whether we work with the adjacency matrix or the graph Laplacian. The paper has two appendices. Appendix~\ref{sec:interp} describes a slight modification of Lagrange barycentric interpolation to compute the companion model and illustrates it with a numerical example.  Appendix~\ref{app:spectralimpulsiverep}, introduces yet another GSP model, the spectral companion model, that is needed to study modulation.
\textbf{Notation}. The following is used throughout: subscripts~$v$, $\textrm{sp}$, $\textrm{comp}$, $\textrm{sp,comp}$ discriminate among similar quantities in different GSP models, namely, vertex, spectral, companion, spectral companion models (below, we ignore the subscripts); sub- or superindex multipl stands for multiplexed;  $\Omega=(\mathcal{A},\mathcal{M},\Phi$ is a GSP model with algebra of filters~$\mathcal{A}$, signal module~$\mathcal{M}$, and bijection~$\Phi$; $s$, $\mathbf{s}$, $\widehat{\mathbf{s}}$, $\mathbf{p}$, $\mathbf{q}$ stand for abstract signal, and different coordinatizations of~$s$ in different models; $A$, $\mathbf{A}$, $\mathbf{C}$ stand for the abstract shift and their coordinatizations in different models; $\mathbf{M}$ is the spectral shift; $B$ refers to basis of~$N$ vectors $\mathbf{b}_n$; $\mathbf{e}_n$, $\boldsymbol{\delta}_n$ are the standard and shifted impulse vectors; $\widehat{\cdot}$ stands for GFT of the quantity; GFT, $\textrm{GFT}^{-1}$, $\mathbf{V}^{-1}$, and $\mathbf{V}$ are the acronyms for the graph Fourier transform and its inverse and their coordinatizations; $\textbf{v}_n$ and $\lambda_n$ are the graph eigenvectors (columns of~$\mathbf{V}$) and graph eigenfrequencies; $\boldsymbol{\Lambda}$ is the diagonal matrix of the eigenvalues; $\boldsymbol{\lambda}$ is the eigenfrequency vector or vector of the graph eigenfrequencies; $\boldsymbol{\lambda}^n$ is the Hadarmard product of~$n$ $\boldsymbol{\lambda}$'s; $\mathcal{V}$ is the Vandermonde matrix; $\mathbf{I}$ and~$\mathbf{1}$ are the identity matrix and the vector of ones; $P(\mathbf{A})$ and $Q(\mathbf{M})$ are polynomial matrices in~$\mathbf{A}$ and~$\mathbf{M}$; $p(x)$ a polynomial in indeterminate~$x$, also the $z$-transform; $\Delta_{\mathbf{A}}(x)$ and $m_{\mathbf{A}}(x)$ are the characteristic and minimum polynomial of~$\mathbf{A}$; $\mathbf{d}^{\textrm{multipl}}$ is multiplexed signal.
\section{Background on GSP}\label{sec:backgroundongsp}
We review graph signal processing (GSP) following \cite{Sandryhaila:13,Sandryhaila:14,Sandryhaila:14big}. Assume a graph $G=(V,E)$ with vertex or node set of finite cardinality $|V|=N$ and edge set~$E$. The graph is defined by the unweighted adjacency matrix~$\mathbf{A}$ of the graph. The entries of~$\mathbf{A}$ are $A_{ij}=1$, if there is graph edge $(i,j)\in E$, $i\neq j$, from source or origin node~$j$ to destination node~$i$, or zero otherwise.\footnote{\label{ftn:adjacencymatrix} This is the standard convention in GSP for adjacency matrices. The computer science literature often reverses the roles of nodes~$i$ and~$j$ in $A_{ij}$ so that the adjacency matrix is $\mathbf{A}^T$, the transpose of~$\mathbf{A}$.}

Let $\left(n_0,\cdots,n_{N-1}\right)$ be a permutation of the~$N$ distinct nodes of~$G$.  A graph signal is the unordered $N$-tuple  $s\!=\!\left(s_{n_0},\!\cdots\!,s_{n_{N-1}}\right)\in\mathbb{C}^V$.
%
The signal set $\mathbb{C}^V$ is a vector space given by the Cartesian product of~$N$ copies of~$\mathbb{C}$.

Algebraic signal processing (ASP) \cite{Pueschel:03a,Pueschel:05e,Pueschel:08a,Pueschel:08b,Pueschel:08c} is an axiomatic approach to signal processing. According to ASP, a graph signal model~$\Omega=(\mathcal{A},\mathcal{M},\Phi)$ is a triplet of:
\begin{inparaenum}[1)]
\item an algebra of filters~$\mathcal{A}$ (a vector space where multiplication of filters is defined);
\item a signal (vector) space~$\mathcal{M}$, $\textrm{dim}\left(\mathcal{M}\right)=\textrm{dim}\left(\mathbb{C}^V\right)$ with the algebraic structure of a module over~$\mathcal{A}$ obtained by defining filtering, i.e., multiplication, of signals $s\!\in\!\mathcal{M}$ by filters $h\!\in\!\mathcal{A}$; and
\item a bijective linear map~$\Phi$ that fixes a basis
\begin{align}\label{eqn:basisB-1}
B{}&=\left\{\mathbf{b}_0,\cdots,\mathbf{b}_{N-1}\right\}
\end{align}
in~$\mathcal{M}$.\footnote{\label{ftn:basis} Recall that, for finite dimensional spaces, the cardinality of a basis of~$\mathcal{M}$ is $|B|=\textrm{dim}(\mathcal{M})$, the basis vectors in~$B$ are nonzero, and they are linearly independent (hence, $\textrm{span}\hspace{-.1cm}-\hspace{-.1cm}\{B\}=\mathcal{M}$, and every signal is represented as a linear combination of the basis vectors).} We assume the basis~$B$ is ordered, which fixes the ordering of the nodes.
\end{inparaenum}
Multiplication of filters and filtering satisfy the usual properties, see \cite{Pueschel:08a} for details.

The diagram in table~\ref{tab:Phidiagram} details the role of~$\Phi$. The set $V^N$ is the set of all permutations of the~$N$ distinct nodes in~$V$.
\begin{table}[htb]
\captionsetup{width=.4\linewidth}
\centering
\begin{tabular}{cccc}
$V^N$&$\xrightarrow{\hspace{.5cm}s\hspace{.5cm}}$&$\mathbb{C}^V$&\\
$\left(n_0,\cdots,n_{N-1}\right)$&$\xmapsto{\hspace{.5cm}\phantom{s}\hspace{.5cm}}$&$s=\left(s_{n_0},\cdots,s_{n_{N-1}}\right)$&\\
  \multirow{3}{*}&&{$\xydownarrow[\begin{gathered}
 \hfill \\
  \hfill \\
  \hfill \\
  \end{gathered}]{}$ }&
 \hspace{-3cm}$\Phi$\\
  \vspace{-.3cm}
  &&&\\
  &&$\mathcal{M}=\mathbb{C}^N$&\\
&&$\mathbf{s}_B=\Phi(s)=\sum_{n=0}^{N-1}s_n\mathbf{b}_n$&
  \end{tabular}
  \caption{$\Phi$ diagram}
   \label{tab:Phidiagram}
  \end{table}
 The top branch shows the abstract signal~$s$ is an unordered $N$-tuple. The vertical branch shows that~$\Phi$ represents the graph signal~$s$ by a linear combination~$\mathbf{s}_B$ of  the basis vectors
\begin{align}\label{eqn:signalrep-1}
s{}&=s_0\mathbf{b}_0+\cdots+s_{N-1}\mathbf{b}_{N-1}\\
\label{eqn:signalrep-2}
{}&=\underbrace{\left[\mathbf{b}_0,\cdots,\mathbf{b}_{N-1}\right]}_{\mathbf{B}}\underbrace{\left[\begin{array}{c}
s_0\\
\vdots\\
s_{N-1}
\end{array}\right]}_{\mathbf{s}_B}.
 \end{align}

 In~\eqref{eqn:signalrep-2}, $\mathbf{B}$ is the matrix whose columns are the basis vectors, and $\mathbf{s}_B$ is the coordinate vector of~$s$ with respect to the basis~$B$.

\begin{remark}[Filters and their coordinate representations]\label{rmk:linearoperator-1}A filter $h\in\mathcal{A}$ is a  linear operator, $h: \mathcal{M}\rightarrow\mathcal{M}$, while~$\mathbf{H}_B$ is its coordinate representation with respect to basis~$B$ in~$\mathcal{M}$. Eigenvalues, trace, and determinant of a linear operator or filter~$h$ are the eigenvalues, trace, and determinant of any matrix $\mathbf{H}_B$ representing the filter~$h$ with respect to any basis~$B$. These are well defined for the linear operator or filter~$h$ because they are invariant to the choice of basis vectors~$B$.
\end{remark}

Of interest in ASP \cite{Pueschel:08a,Pueschel:08b} and in GSP \cite{Sandryhaila:13} are commutative filters, i.e., when~$\mathcal{A}$ is a commutative algebra. In this case, $\mathcal{A}$ is generated by a single filter, the shift filter~$A$. Following \cite{Sandryhaila:13}, the shift~$A$ in GSP is the linear filter such that sample~$n$ of~$As$ is
\begin{align}\label{eqn:shiftvertexmodelA}
(A s)_{n}{}&=\sum_{j\in\eta_n} s_{j}
\end{align}
where $\eta_{n}$ is the neighborhood of node~$n$ in graph~$G$.

Filters in GSP are chosen to commute with the shift~$A$, i.e., are linear shift invariant (LSI) filters. Under broad conditions  \cite{Sandryhaila:13}, filters are LSI if and only if (iff) they are polynomial filters $P(A)$ of the shift with degree at most $N-1$.

The graph Fourier transform $\textrm{GFT}$ is defined \cite{Pueschel:08a,Sandryhaila:13}  as the transform that diagonalizes the shift.

\textbf{In summary}, choice of an (ordered) basis~$B$ and shift~$A$ define a (commutative) GSP model~$\Omega=(\mathcal{A},\mathcal{M},\Phi)$ on a graph $G=(V,E)$, where $\mathcal{A}=\left\{P(A)\right\}$ is the set of polynomial filters, $\mathcal{M}=\mathbb{C}^N$, and $\Phi$ is defined by~$B$. The graph Fourier transform (GFT) and other GSP concepts follow \cite{Sandryhaila:13,Sandryhaila:14}.

We state formally here for easy reference the main assumptions considered in this paper.

\begin{assumption}[Connected graph]\label{ass:connectedgraph}
The graph~$G$ is simple, and strongly connected.
\end{assumption}
A graph is simple if there are no self-loops, nor multiedges. For simple graphs the diagonal entries of the adjacency matrix are zero. Note that assumption~\ref{ass:connectedgraph} does not further restrict the graph~$G$ that can be undirected, directed, or mixed.

\begin{assumption}[Diagonalizable adjacency matrix~$\mathbf{A}$]\label{ass:diagonalizableA}
The adjacency matrix~$\mathbf{A}$ of the graph~$G$ is diagonalizable.
\end{assumption}

\begin{assumption}[Adjacency matrix~$\mathbf{A}$ with distinct eigenvalues]\label{ass:Adistincteigenvalues}
The adjacency matrix~$\mathbf{A}$ of the graph~$G$ has distinct eigenvalues.
\end{assumption}
Assumption~\ref{ass:Adistincteigenvalues} is a sufficient condition for assumption~\ref{ass:diagonalizableA}. If the graph is undirected, the shift is symmetric and assumption~\ref{ass:diagonalizableA} is satisfied. These assumptions are adopted here for ease of exposition, see \cite{Sandryhaila:13,derimoura-2017} on how to handle nondiagonalizable~$\mathbf{A}$.
\section{The GSP Vertex and Spectral Models}\label{sec:vertexspectralmodels}
GSP models share~$\mathcal{A}$ and~$\mathcal{M}$. They are distinguished by~$\Phi$ that determines the basis~$B$ in~$\mathcal{M}$.
We illustrate two GSP models, the GSP vertex model $\Omega_{\textrm{v}}$ and the GSP spectral model  $\Omega_{\textrm{sp}}$.
\subsection{GSP Vertex Model $\Omega_{\textrm{v}}$}\label{subsec:gspvertexmodel}
This is the natural model introduced in \cite{Sandryhaila:13} and further developed in \cite{Sandryhaila:14,Sandryhaila:14big}, see also \cite{ortegafrossardkovacevicmouravandergheynst-2018}.

\textbf{Basis}. The basis  $B_{\textrm{v}}=\left\{\mathbf{e}_n\right\}_{0\leq n\leq N-1}$ is the set of~$N$ nonzero, linearly independent standard vectors (hence a basis), where $\mathbf{e}_n$ is a vector of zeros, except for entry~$n$ that is a one. The standard signal representation is
\begin{align}\label{eqn:graphsignals-naturalrep}
\mathbf{s}_{\textrm{v}}{}&=\underbrace{\left[\begin{array}{c}
1\\
0\\
\vdots\\
0\\
0
\end{array}\right]}_{\mathbf{e}_0}s_0+\cdots +\underbrace{\left[\begin{array}{c}
0\\
0\\
\vdots\\
0\\
1
\end{array}\right]}_{\mathbf{e}_{N-1}}s_{N-1}\\
\label{eqn:graphsignals}
{}&=\underbrace{\left[\begin{array}{cccc}
\mathbf{e}_0\,\mathbf{e}_1\cdots \mathbf{e}_{N-1}
\end{array}\right]}_{\mathbf{B}_v=\mathbf{I}_N}\underbrace{\left[\begin{array}{c}
s_0\\
\vdots\\
s_{N-1}
\end{array}\right]}_{\mathbf{s}_{\textrm{v}}}
\end{align}
The standard representation is also called the vertex or the Euclidean representation. The vector $\mathbf{s}_{\textrm{v}}$ is the coordinate vector of the graph signal~$s$ with respect to the standard basis $B_{\textrm{v}}$.  We note that $\mathbf{e}_n$ are the coordinate vector representations with respect to $B_{v}$ of  graph signals that are zero everywhere, except at node~$n$ where they are one.

\textbf{Shift}.  With respect to the basis $B_{\textrm{v}}$, it is easy to see that the shift~$A$ in~\eqref{eqn:shiftvertexmodelA} is represented by the adjacency matrix~$\mathbf{A}$ of the graph~$G$.

The graph associated with the vertex GSP model is of course the original graph~$G$. The nodes of the graph are associated with the basis vectors $\mathbf{e_n}$ of the basis~$B$.

\begin{remark}\label{rmk:linearoperator-2}
Since the matrix $\mathbf{B}_v$ whose columns are the standard basis vectors is the identity~$\mathbf{I}_N$, see equation~\eqref{eqn:graphsignals}, and by remark~\ref{rmk:linearoperator-1}, eigenvalues, trace, and determinant of a linear operator are invariant to the choice of basis~$B$, in the sequel, unless otherwise stated, we will work with the standard coordinate vector and matrix representations, even when dealing with the abstract concept of a  signal~$s$, shift~$A$, or filter~$h$.
\end{remark}

\begin{remark}[Dropping index~$v$]\label{rmk:droppingindexv}
    In the sequel, we often drop the subindex~$v$ when referring to coordinatizations with respect to the standard basis~$B_v$. For example, for short, we may refer to $\mathbf{s}_v$  simply as~$\mathbf{s}$ and as the graph signal.
\end{remark}

\textbf{Graph Fourier transform (GFT)}. The coordinate representation  $\mathbf{V}^{-1}$ of the $\textrm{GFT}_{\textrm{v}}$, under assumption~\ref{ass:diagonalizableA}, follows \cite{Sandryhaila:13} by  diagonalizing~$\mathbf{A}$
\begin{align}\label{eqn:GFT-1}
    \mathbf{A}{}&=\textrm{GFT}_{\textrm{v}}^{-1}\cdot\boldsymbol{\Lambda}^\cdot\textrm{GFT}_{\textrm{v}}\\
 \label{eqn:GFT-2}   {}&=\mathbf{V}\boldsymbol{\Lambda}\mathbf{V}^{-1}.
\end{align}
Equations~\eqref{eqn:GFT-1} and~\eqref{eqn:GFT-2} show that, as referred to before, $\mathbf{V}^{-1}$ is the graph Fourier transform $\textrm{GFT}_{\textrm{v}}$ in the vertex model, the columns of~$\mathbf{V}$ and the diagonal entries of the diagonal matrix~$\boldsymbol{\Lambda}$ are the eigenvectors $\mathbf{v}_n$ and the eigenvalues $\lambda_n$ of~$\mathbf{A}$. The eigenvectors $\left\{\mathbf{v}_n\right\}_{0\leq n\leq N-1}$ and the eigenvalues $\left\{\lambda_n\right\}_{0\leq n\leq N-1}$ are called the graph or Fourier modes and the graph frequencies of the GSP vertex model $\Omega_{\textrm{v}}$. The matrix~$\mathbf{V}$ will also be referred to as the Fourier basis.

The coordinate vector~$\widehat{\mathbf{s}}$ of the $\!\textrm{GFT}_{\textrm{v}}\!$ of the graph signal~$\mathbf{s}$ is
\begin{align}\label{eqn:coortdinatevectorwidehats}
\widehat{\mathbf{s}}{}&=\textrm{GFT}_v\cdot\mathbf{s}=\mathbf{V}^{-1}\cdot \mathbf{s}.
\end{align}
\subsection{GSP Spectral Model $\Omega_{\textrm{sp}}$}\label{subsec:gspspmodel}
We consider a second GSP model. By assumption~\ref{ass:diagonalizableA}, the~$N$ columns, $\left\{\mathbf{y}_n\right\}_{0\leq n\leq N-1}$, of $\mathbf{V}^{-1}$, the $\textrm{GFT}_{\textrm{v}}$  of the GSP vertex model $\Omega_{\textrm{v}}$, are linearly independent and form a basis of $\mathcal{M}=\mathbb{C}^N$.
By choosing $B_{\textrm{sp}}=\left\{\mathbf{y}_n\right\}_{0\leq n\leq N-1}$ as basis of~$\mathcal{M}$, we obtain the GSP spectral model $\Omega_{\textrm{sp}}$, the basis is . The coordinate representation $\mathbf{s}_{\textrm{sp}}$ of the signal~$\mathbf{s}$ with respect to $B_{\textrm{sp}}$ is
\begin{align}\label{eqn:s-spcoordinaterep-1}
\mathbf{s}_{\textrm{sp}}{}&=\sum_{m=0}^{N-1}\mathbf{y}_m s_m
=\left[\mathbf{y}_0\,\cdots\,\mathbf{y}_{N-1}\right]\mathbf{s}=\mathbf{V}^{-1}\mathbf{s}\\
\label{eqn:s-spcoordinaterep-3}
{}&=\widehat{\mathbf{s}}.
\end{align}
In other words, $\mathbf{s}_{\textrm{sp}}=\widehat{\mathbf{s}}$, the GFT of~$\mathbf{s}$.

\textbf{Shift}. To determine the appropriate coordinate representation~$A_{\textrm{sp}}$ of the shift in the GSP spectral model $\Omega_{\textrm{sp}}$, we first determine the graph Fourier transform $\textrm{GFT}_{\textrm{sp}}$ in the GSP model $\Omega_{\textrm{sp}}$. From~\eqref{eqn:s-spcoordinaterep-1}-\eqref{eqn:s-spcoordinaterep-3}, we have
\begin{align}\label{eqn:gftspmodel-1}
    \mathbf{s}{}&=\textrm{GFT}_{\textrm{sp}}\cdot \widehat{\mathbf{s}}\\
    \label{eqn:gftgspspmodel-2}
{}&=\mathbf{V}\cdot\widehat{\mathbf{s}}.
\end{align}
In other words, the $\textrm{GFT}_{\textrm{sp}}$ in the GSP $\Omega_{\textrm{sp}}$ is~$\mathbf{V}$, the inverse graph Fourier transform $\textrm{GFT}_{\textrm{v}}^{-1}$ of the GSP vertex model $\Omega_{\textrm{v}}$. We see that in the $\Omega_{\textrm{sp}}$ and $\Omega_{\textrm{v}}$ models the graph Fourier transforms are the inverses of each other. References \cite{shimoura-asilomar2019,shimoura-2021} show that the coordinate matrix~$\mathbf{M}$ of the shift $A_{\textrm{sp}}$ in the GSP spectral model $\Omega_{\textrm{sp}}$ is then
\begin{align}\label{eqn:shiftgspspmodel-1}
\mathbf{M}{}&=\textrm{GFT}_{\textrm{sp}}^{-1}\cdot\boldsymbol{\Lambda}^*\cdot\textrm{GFT}_{\textrm{sp}}\\
\label{eqn:shiftgspspmodel-2}
{}&=\mathbf{V}^{-1}\cdot\boldsymbol{\Lambda}^*\cdot\mathbf{V},
\end{align}
where~$*$ stands for complex conjugate, see \cite{shimoura-asilomar2019,shimoura-2021} for details.

Further, interpreting the spectral shift~$\mathbf{M}$ as an adjacency matrix of a graph, we obtain the graph $G_{\textrm{sp}}=\left(V,E_{\textrm{sp}}\right)$ associated with the GSP spectral model.

In the GSP spectral model $\Omega_{\textrm{sp}}$, graph filters are polynomials $Q(\mathbf{M})$ of the spectral shift~$\mathbf{M}$, not of the vertex shift~$\mathbf{A}$. In \cite{shimoura-2021}, the spectral model $\Omega_{\textrm{sp}}$ and the spectral shift~$\mathbf{M}$ are used to establish the relation between GSP sampling in the vertex and graph spectral domains, a dual relationship that mimics the four steps of downsampling, decimation, upsampling, and reconstruction by LSI filtering in Shannon sampling of band-limited time signals. For example, \cite{shimoura-2021} shows that, as with Shannon sampling of time signals,  the graph spectrum of the downsampled version of a bandlimited graph signal is a replication of the graph spectrum of the original signal, except that, in contrast with \cite{tanaka-2018}, the replicated replicas are  distorted replicas of the original spectrum.
\section{The GSP Companion Model $\Omega_{\textrm{comp}}$}\label{sec:gspcompanionmodel}
Section~\ref{subsec:gspvertexmodel} presented the GSP vertex model $\Omega_{\textrm{v}}$ that is the standard model adopted by the original papers \cite{Sandryhaila:13,Sandryhaila:14,Sandryhaila:14big}, see also \cite{ShumanNFOV:13} and \cite{ortegafrossardkovacevicmouravandergheynst-2018}, as well as by the majority of GSP papers. This is natural, since graph signals are commonly specified by their values at the nodes of the underlying graph~$G$.

Section~\ref{subsec:gspspmodel} presented a second GSP model, the spectral model $\Omega_{\textrm{sp}}$, whose spectral shift~$\mathbf{M}$ was introduced in \cite{shimoura-asilomar2019} and that was further considered and applied to develop graph sampling in \cite{shimoura-2021}.

This section considers a third model, the GSP companion model $\Omega_{\textrm{comp}}$. It will help addressing the following issues.

\textbf{Ubi sunt\footnote{\label{lab:ubisunt}Latin expression ``where are.''} the boundary conditions (bc)?}
When we shift the samples at the \textit{boundary} of a finite duration signal we run into the issue of \textit{boundary conditions} (bc), also known as signal \textit{continuation}. The bc specify how the shifted \textit{boundary} samples are obtained from the~$N$ known samples of the signal. There are several ways this is accomplished in practice, periodization and extending the signals with zeros are but two common practices. In ASP,  \cite{Pueschel:05e,Pueschel:08a,Pueschel:08b} specify explicitly the boundary conditions for the time model and the 1D-space model. Surprisingly, even though graphs are assumed of finite size (finite number of vertices), no one in the GSP literature that we are aware of, not the original papers \cite{Sandryhaila:13,Sandryhaila:14,Sandryhaila:14big}, see also \cite{ShumanNFOV:13}, and also \cite{ortegafrossardkovacevicmouravandergheynst-2018}, nor the many follow-up papers, has specified explicitly the boundary conditions~(bc). It is natural to ask where are the bc in the GSP models. The GSP companion shift makes them explicit.

\textbf{GSP as a replication of DSP: The step~1-step~2 companion model method (S$^2$CM$^2$)}. Reference \cite{Sandryhaila:13} introduced GSP as an extension of DSP, by first capturing DSP in a graph framework. Many other authors use the intuition provided by DSP to derive new concepts in GSP. Because these approaches work directly in the GSP vertex model $\Omega_{\textrm{v}}$, they have required significant effort on the part of the authors, sometimes, with limited success. It is reasonable to ask if there is a GSP model that replicates reasonably close the DSP time model. We show here that the GSP companion model achieves this,  and we use it to introduce a systematic step~1-step~2 \textit{companion} model method (S$^2$CM$^2$) to extend GSP:
\begin{inparaenum}[1)]
\item replicate the DSP concept in the GSP companion model $\Omega_{\textrm{comp}}$, and then
\item transform it back to the GSP vertex model $\Omega_{\textrm{v}}$.
\end{inparaenum}

To develop $\Omega_{\textrm{comp}}$, we introduce a new basis, the companion basis $B_{\textrm{comp}}$ and then a new shift, $A_{\textrm{comp}}$. From these, all other usual GSP concepts follow, including the graph Fourier transform $\textrm{GFT}_{\textrm{comp}}$, the companion eigenmodes and eigenfrequencies, and the companion graph $G_{\textrm{comp}}$. We illustrate in subsequent sections how S$^2$CM$^2$ extends GSP with new concepts without much effort.

\subsection{The Companion Basis $B_{\textrm{comp}}$}\label{subsec:companionbasis}
\begin{figure}[htb!]
\centering	\includegraphics[width=4cm, keepaspectratio]{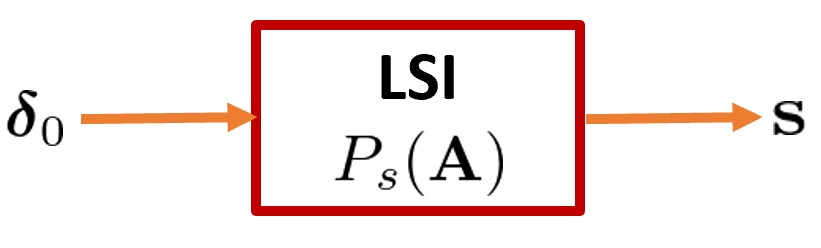}
	\caption{Graph signal~$\mathbf{s}$ as impulse response of LSI filter $P_s(\mathbf{A})$}
\label{fig:sasimpulseresp}
\end{figure}

The companion basis $B_{\textrm{comp}}$ that we introduce here interprets the graph signal~$\mathbf{s}$ as the impulse response of a graph LSI filter $P_{s}(\mathbf{A})$, see figure~\ref{fig:sasimpulseresp}.

To develop $B_{\textrm{comp}}$, we first consider the graph impulse $\delta_0$ and its shifts $\delta_n$.
\subsubsection{Graph impulse}\label{subsubsec:graphimpulse} We can consider two alternatives to  define $\delta_0$:
\begin{inparaenum}[1)]
\item\label{inp:impulsivedelta} impulsive in the vertex domain: choose $\delta_0$ to be the standard graph signal $e_0$ that is~1 in a vertex labeled~0 and~0 at all other nodes; or
\item \label{inp:flatdelta} choose $\delta_0$ to have flat graph spectrum. \end{inparaenum}
In DSP, these two alternatives define the same signal, but in the GSP vertex model $\Omega_{v}$ this is not the case. Alternative~\ref{inp:impulsivedelta} presents the difficulty of which vertex should be chosen as vertex~0, while alternative~\ref{inp:flatdelta} treats equally every eigenfrequency. We choose alternative~\ref{inp:flatdelta}. We obtain the coordinate vector $\boldsymbol{\delta}_0$ representation of $\delta_0$ in $\Omega_{v}$ from its graph spectrum given by
\begin{align}\label{eqn:flatdelta-1}
\widehat{\boldsymbol{\delta}}_0{}&=\frac{1}{\sqrt{N}}\mathbf{1}
\end{align}
where~$\mathbf{1}$ is the vector of ones and the factor $\frac{1}{\sqrt{N}}$ normalizes $\widehat{\boldsymbol{\delta}}_0$ to norm~1. The coordinate vector $\boldsymbol{\delta}_0$ representation of $\delta_0$ in the vertex domain is then
\begin{align}\label{eqn:flatdelta-2}
\boldsymbol{\delta}_0{}&=\textrm{GFT}_{\textrm{v}}^{-1}\cdot
\widehat{\boldsymbol{\delta}}_0= \mathbf{V}\cdot \frac{1}{\sqrt{N}}\mathbf{1}.
\end{align}
\subsubsection{Shifted graph impulse}\label{subsubsec:shiftedimpulse}
We consider the shifted graph impulse and its graph spectrum in the GSP vertex model $\Omega_{\textrm{v}}$. We shift a graph signal by multiplying it by the shift~$\mathbf{A}$. For example, using the eigendecomposition~\eqref{eqn:GFT-2} of~$\mathbf{A}$ and the definition~\eqref{eqn:flatdelta-1} of $\widehat{\boldsymbol{\delta}}_0$ we obtain successively
\begin{align}\label{eqn:shiftedelta1-1}
\boldsymbol{\delta}_1{}&=\mathbf{A}\cdot \boldsymbol{\delta}_0
=\mathbf{V}\boldsymbol{\Lambda}\mathbf{V}^{-1}\boldsymbol{\delta}_0
=\mathbf{V}\boldsymbol{\Lambda}\frac{1}{\sqrt{N}}\mathbf{1}\\
\label{eqn:shiftedelta1-4}
{}&=\mathbf{V}\frac{1}{\sqrt{N}}\boldsymbol{\lambda}
\xrightarrow{\textrm{GFT}_{\textrm{v}}}
\widehat{\boldsymbol{\delta}}_1=\frac{1}{\sqrt{N}}\boldsymbol{\lambda}
\end{align}
where
\begin{align}\label{eqn:shiftedelta1-5}
    \boldsymbol{\lambda}{}&=\boldsymbol{\Lambda}\mathbf{1}=\left[\begin{array}{c}
    \lambda_0\\
    \vdots\\
    \lambda_{N-1}
    \end{array}\right]
\end{align}
is the vector of the eigenfrequencies of the graph~$G$, i.e., the vector of the eigenvalues of its adjacency matrix~$\mathbf{A}$. We refer to~$\boldsymbol{\lambda}$ as the \textit{eigenfrequency} vector. Equation~\eqref{eqn:shiftedelta1-4} shows that the graph Fourier transform $\widehat{\boldsymbol{\delta}}_1$ of the shifted impulse $\boldsymbol{\delta}_1$ is the (normalized) eigenfrequency vector $\boldsymbol{\lambda}$.

Repeating the steps in equations~\eqref{eqn:shiftedelta1-1}--\eqref{eqn:shiftedelta1-4} $n$~times, we obtain the $n$th-shifted impulse and its spectrum in the GSP vertex model to be
\begin{align}\label{eqn:shiftedelta1-6}
\hspace{-.4cm}\boldsymbol{\delta}_n{}&=\mathbf{A}^n\boldsymbol{\delta}_0=\textrm{GFT}_{\textrm{v}}^{-1}\frac{1}{\sqrt{N}}\boldsymbol{\lambda}^n\!=\!\mathbf{V}\frac{1}{\sqrt{N}}\boldsymbol{\lambda}^n\!\xrightarrow{\textrm{GFT}_{\textrm{v}}} \widehat{\boldsymbol{\delta}}_n\!=\!\frac{1}{\sqrt{N}}\boldsymbol{\lambda}^n
\end{align}
where, using the Hadamard or component wise product~$\circ$,
\begin{align}\label{eqn:hadamard}
\boldsymbol{\lambda}^n{}&=\underbrace{\boldsymbol{\lambda}\circ\cdots\circ\boldsymbol{\lambda}}_{n\textrm{   times   }}
\end{align}

\subsubsection{The companion basis $B_{\textrm{comp}}$}\label{subsubsec:companionbasis}
We now return to the companion basis. We take it as the impulse and its $N-1$ shifts
\begin{align}\label{eqn:companionbasis}
B_{\textrm{comp}}{}&=\left\{\boldsymbol{\delta}_0,\mathbf{A}\boldsymbol{\delta}_0,\cdots,\mathbf{A}^{N-1}\boldsymbol{\delta}_0\right\}\\
\label{eqn:companionbasis-1}
{}&=\left\{\boldsymbol{\delta}_0,\boldsymbol{\delta}_1,\cdots,\boldsymbol{\delta}_{N-1}\right\}
\end{align}
If we take the graph Fourier transform $\textrm{GFT}=\mathbf{V}^{-1}$ of the impulse and its shifts in $B_{\textrm{comp}}$, we get the set
\begin{align}\label{eqn:companionbasis-2}
\widehat{B}_{\textrm{comp}}{}&=\left\{\widehat{\boldsymbol{\delta}}_0=\frac{1}{\sqrt{N}}\mathbf{1},\widehat{\boldsymbol{\delta}}_{1}=\frac{1}{\sqrt{N}}\boldsymbol{\lambda},\cdots,\widehat{\boldsymbol{\delta}}_{N-1}=\frac{1}{\sqrt{N}}\boldsymbol{\lambda}^{N-1}\right\}
\end{align}
Under assumption~\ref{ass:Adistincteigenvalues}, it is well known that the~$N$ vectors in the set $\widehat{B}_{\textrm{comp}}$ are nonzero and linearly independent (see also discussion on Vandermonde matrices below and equation~\eqref{eqn:detvandermonde}). Since  $\textrm{GFT}_{\textrm{v}}^{-1}=\mathbf{V}$ is full rank, the~$N$ vectors in $B_{\textrm{comp}}$ are linearly independent and $B_{\textrm{comp}}$ is a basis.

\subsubsection{The companion representation $\mathbf{p}_{\textrm{comp}}$}\label{subsubsec:companionrep}
We represent the graph signal with respect to the companion basis $B_{\textrm{comp}}$. Get
\begin{align}\label{eqn:companionrep-1}
\mathbf{s}{}&=p_0\boldsymbol{\delta}_0+p_1\boldsymbol{\delta}_1+\cdots+p_{N-1}\boldsymbol{\delta}_{N-1}\\
\label{eqn:companionrep-2}
{}&=\underbrace{\left[\boldsymbol{\delta}_0\,\boldsymbol{\delta}_1\cdots \boldsymbol{\delta}_{N-1}\right]}_{\mathbf{B}_{\textrm{comp}}}\underbrace{\left[\begin{array}{c}
p_0\\
p_1\\
\vdots\\
p_{N-1}
\end{array}\right]}_{\mathbf{p}_{\textrm{comp}}}
\end{align}
The vector $\mathbf{p}_{\textrm{comp}}$ is the coordinate vector of the graph signal~$\mathbf{s}$ with respect to the companion basis $B_{\textrm{comp}}$. The columns of the matrix $\mathbf{B}_{\textrm{comp}}$ in~\eqref{eqn:companionrep-2} are the coordinate vectors of the impulse and its $N-1$ shifts.

\textbf{Graph signal as impulse response of LSI filter $P_s(\mathbf{A})$}.
Replacing in~\eqref{eqn:companionrep-1} the shifted impulses $\boldsymbol{\delta}_{n}$ by $\mathbf{A}^n\boldsymbol{\delta}_0$, and factoring out the impulse $\boldsymbol{\delta}_0$, obtain
\begin{align}\label{eqn:companionrep-3}
\mathbf{s}{}&=\underbrace{\left\{p_0\mathbf{I} +p_1\mathbf{A}+\cdots+p_{N-1}\mathbf{A}^{N-1}\right\}}_{P_s(\mathbf{A})}\boldsymbol{\delta}_0
\end{align}
 Equation~\eqref{eqn:companionrep-3} interprets the graph signal as the impulse response of the LSI filter $P_s(\mathbf{A})$, see figure~\ref{fig:sasimpulseresp}. The vector of coefficients of $P_s(\mathbf{A})$ is the coordinate vector $\mathbf{p}_{\textrm{comp}}$.

\textbf{Determination of $\mathbf{p}_{\textrm{comp}}$}. Given the signal~$\mathbf{s}$, the coordinate vector $\mathbf{p}_{\textrm{comp}}$ can be determined by solving the linear system
\begin{align}\label{eqn:companionrep-4}
\mathbf{B}_{\textrm{comp}} \cdot \mathbf{p}_{\textrm{comp}}{}&=\mathbf{s}.
\end{align}
Since the impulse and its $N-1$ shifts are linearly independent, $\mathbf{B}_{\textrm{comp}}$ is full rank,
equation~\eqref{eqn:companionrep-4} has a unique solution, and $\mathbf{p}_{\textrm{comp}}$ is unique.

We provide an alternative method to determine $\mathbf{p}_{\textrm{comp}}$. Replacing~$\mathbf{A}$ in~\eqref{eqn:companionrep-3} by its diagonalization in~\eqref{eqn:GFT-2}, and exchanging the left-hand-side with the right-hand-side, get
\begin{align}\label{eqn:companionrep-5}
\left\{p_0\mathbf{I} +p_1\mathbf{V}\boldsymbol{\Lambda}\mathbf{V}^{-1}+\cdots+p_{N-1}\mathbf{V}\boldsymbol{\Lambda}^{N-1}\mathbf{V}^{-1}\right\}\boldsymbol{\delta}_0{}&=\mathbf{s}
\end{align}
Factoring~$\mathbf{V}$ to the left and $\mathbf{V}^{-1}$ to the right, obtain
\begin{align}\label{eqn:companionrep-6}
\mathbf{V}\underbrace{\left\{p_0\mathbf{I} +p_1\boldsymbol{\Lambda}+\cdots+p_{N-1}\boldsymbol{\Lambda}^{N-1}\right\}}_{P_s(\boldsymbol{\Lambda)}}\underbrace{\mathbf{V}^{-1}\boldsymbol{\delta}_0}_{\widehat{\boldsymbol{\delta}}_0}{}&=\mathbf{s}
\end{align}
The matrix $P_s(\boldsymbol{\Lambda})$ is the diagonal matrix
\begin{align}\label{eqn:Pslambda-1}
P_s(\boldsymbol{\Lambda}){}&=\textrm{diag}\left[P_s(\boldsymbol{\lambda}_0),P_s(\boldsymbol{\lambda}_1),\cdots,P_s(\boldsymbol{\lambda}_{N-1})\right]\\
\label{eqn:Pslambda-2}
{}&=\left[\begin{array}{cccc}
P_s(\boldsymbol{\lambda}_0)&&&\\
& P_s(\boldsymbol{\lambda}_1)&&\\
&&\ddots\\
&&&P_s(\boldsymbol{\lambda}_{N-1})
\end{array}\right]
\end{align}
where $P_s(\boldsymbol{\lambda}_n)$ is the polynomial $P_s(\mathbf{A})$ evaluated at the eigenfrequency $\lambda_n$.

Replacing in equation~\eqref{eqn:companionrep-6}  $\widehat{\boldsymbol{\delta}}_0$ by its definition in~\eqref{eqn:shiftedelta1-6}, taking the graph Fourier transform to both sides of~\eqref{eqn:companionrep-6}, and using $P_s(\boldsymbol{\Lambda})$ given in~\eqref{eqn:Pslambda-1} obtain
\begin{align}\label{eqn:vandermonde-1}
\hspace{-.2cm}\left[\!\!\begin{array}{ccc}
P_s(\boldsymbol{\lambda}_0)&&\\
&\ddots\\
&&P_s(\boldsymbol{\lambda}_{N-1})
\end{array}\!\!\right]\!\frac{1}{\sqrt{N}}\mathbf{1}\!=&\!\\
\label{eqn:vandermonde-2}
\frac{1}{\sqrt{N}}\!\left[\!\!\!\begin{array}{c}
P_s(\boldsymbol{\lambda}_0)\\
\vdots\\
P_s(\boldsymbol{\lambda}_{N-1})
\end{array}\!\!\!\right]\!\!=\!\widehat{\mathbf{s}}
\end{align}
Factoring out $\mathbf{p}_{\textrm{comp}}$ on the left-hand-side of equation~\eqref{eqn:vandermonde-2}, get
\begin{align}\label{eqn:vandermonde-3}
\left\{\begin{array}{llll}
&\frac{1}{\sqrt{N}}\underbrace{\left[\begin{array}{cccc}
1&\lambda_0&\cdots&\lambda_0^{N-1}\\
\vdots&\vdots&\vdots&\vdots\\
1&\lambda_{N-1}&\cdots&\lambda_{N-1}^{N-1}
\end{array}\right]}_{\mathcal{V}}&\mathbf{p}_{\textrm{comp}}{}&=\\
&\hspace{.32cm}\frac{1}{\sqrt{N}}\overbrace{\left[\begin{array}{cccc}
\mathbf{1}&\boldsymbol{\lambda}&\cdots&\boldsymbol{\lambda}^{N-1}
\end{array}\right]}_{}&\mathbf{p}_{\textrm{comp}}{}&=\widehat{\mathbf{s}}
\end{array}\right.
\end{align}
The matrix~$\mathcal{V}$ in~\eqref{eqn:vandermonde-3} is a Vandermonde matrix with determinant
\begin{align}\label{eqn:detvandermonde}
|\mathcal{V}|{}&=\Pi_{i,j,i\neq j}\left(\lambda_i-\lambda_j\right)
\end{align}
that under assumption~\ref{ass:Adistincteigenvalues} is full rank and invertible.

For future reference, we rewrite equations~\eqref{eqn:companionrep-4} and~\eqref{eqn:vandermonde-3} as
\begin{align}\label{eqn:vandermonde-5}
\frac{1}{\sqrt{N}}\mathbf{V}\mathcal{V}\cdot \mathbf{p}_{\textrm{comp}}{}&=\mathbf{s}\\
\label{eqn:vandermonde-6}
\frac{1}{\sqrt{N}}\mathcal{V}\cdot\mathbf{p}_{\textrm{comp}}{}&=\widehat{\mathbf{s}}
\end{align}
Equations~\eqref{eqn:vandermonde-5} (or equivalently~\eqref{eqn:companionrep-4}) and~\eqref{eqn:vandermonde-6} can be used to obtain $\mathbf{p}_{\textrm{comp}}$ from $\mathbf{s}$ or from~$\widehat{\mathbf{s}}$. We can consider regularized methods to solve either of the systems in~\eqref{eqn:vandermonde-5} (or equivalently~\eqref{eqn:companionrep-4}) or~\eqref{eqn:vandermonde-6} to determine $\mathbf{p}_{\textrm{comp}}$ from~$\mathbf{s}$ or~$\widehat{\mathbf{s}}$. As an alternative, appendix~\ref{sec:interp} describes  Lagrange barycentric interpolation to solve these equations without inversion of the system matrices. 
%
 \begin{figure}[htb!]
\centering	\includegraphics[width=6cm, keepaspectratio]{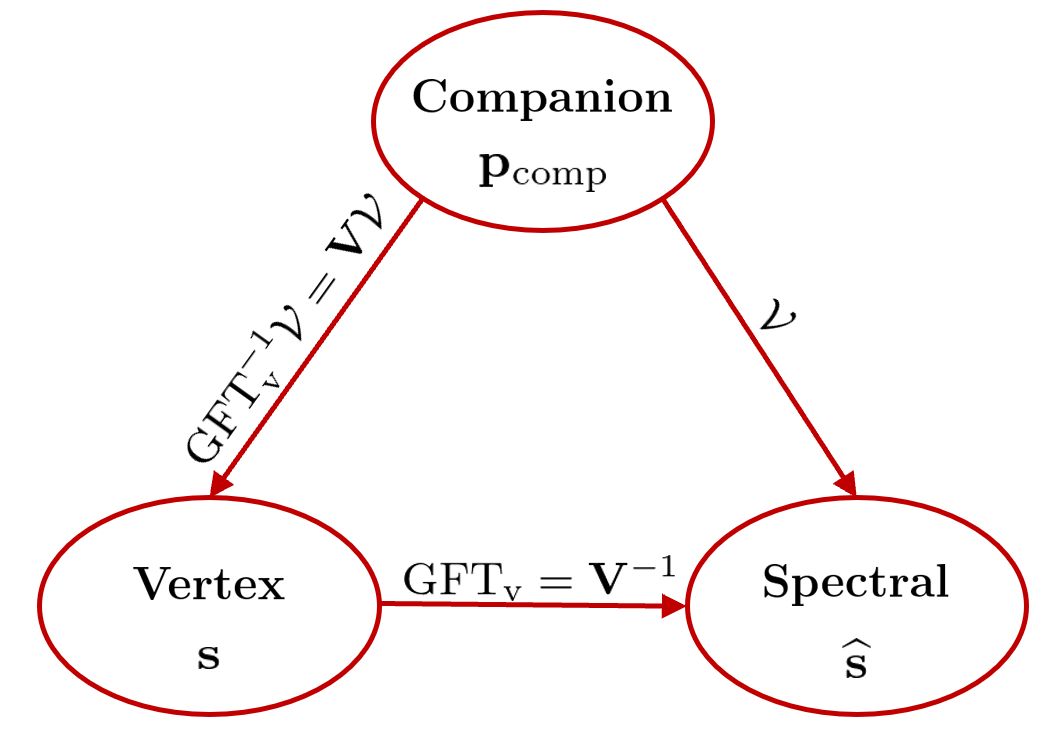}
	\caption{Vertex, spectral, and companion GSP domains}
\label{fig:3signaldomains-1}
\end{figure}
Figure~\ref{fig:3signaldomains-1} illustrates the relations between the three signal representations and the transforms between them given by~\eqref{eqn:vandermonde-5} and~\eqref{eqn:vandermonde-6}. We will also refer to $\textrm{GFT}^{-1}_{\textrm{v}}\boldsymbol{\mathcal{V}}=\mathbf{V}\boldsymbol{\mathcal{V}}$ and $\boldsymbol{\mathcal{V}}$ that recover~$\mathbf{s}$ and~$\widehat{\mathbf{s}}$ from~$\mathbf{p}_{\textrm{comp}}$ as confusion factors.

\begin{example}[Impulsive signals in different domains]\label{exp:impulsivesignals}
This example illustrates for the three domains of figure~\ref{fig:3signaldomains-1} two signals: one is impulsive in the vertex domain, and the other is flat in the spectral domain. We choose as graph~$G$ the \textit{directed} simple graph of 35 vertices shown in
figure~\ref{fig:2023-12-21-eg351}. The characteristic polynomial of~$G$ is
$\Delta_\mathbf{A}(x)=x^{35}-x^{31}-x^{30}-2 x^{29}-x^{28}-3 x^{27}-x^{26}-3 x^{25}+2 x^{24}+x^{23}+4 x^{21}-x^{20}+5 x^{19}-3 x^{18}+x^{17}-6 x^{16}+2 x^{14}-3 x^{13}+6 x^{12}-4 x^{11}-3 x^{10}-8 x^9+6 x^8+x^7-4 x^6-2 x^5-x^4+4 x^3-x^2+x-1$.
 \begin{figure}[htb!]
\centering	\includegraphics[width=8.5cm, keepaspectratio]{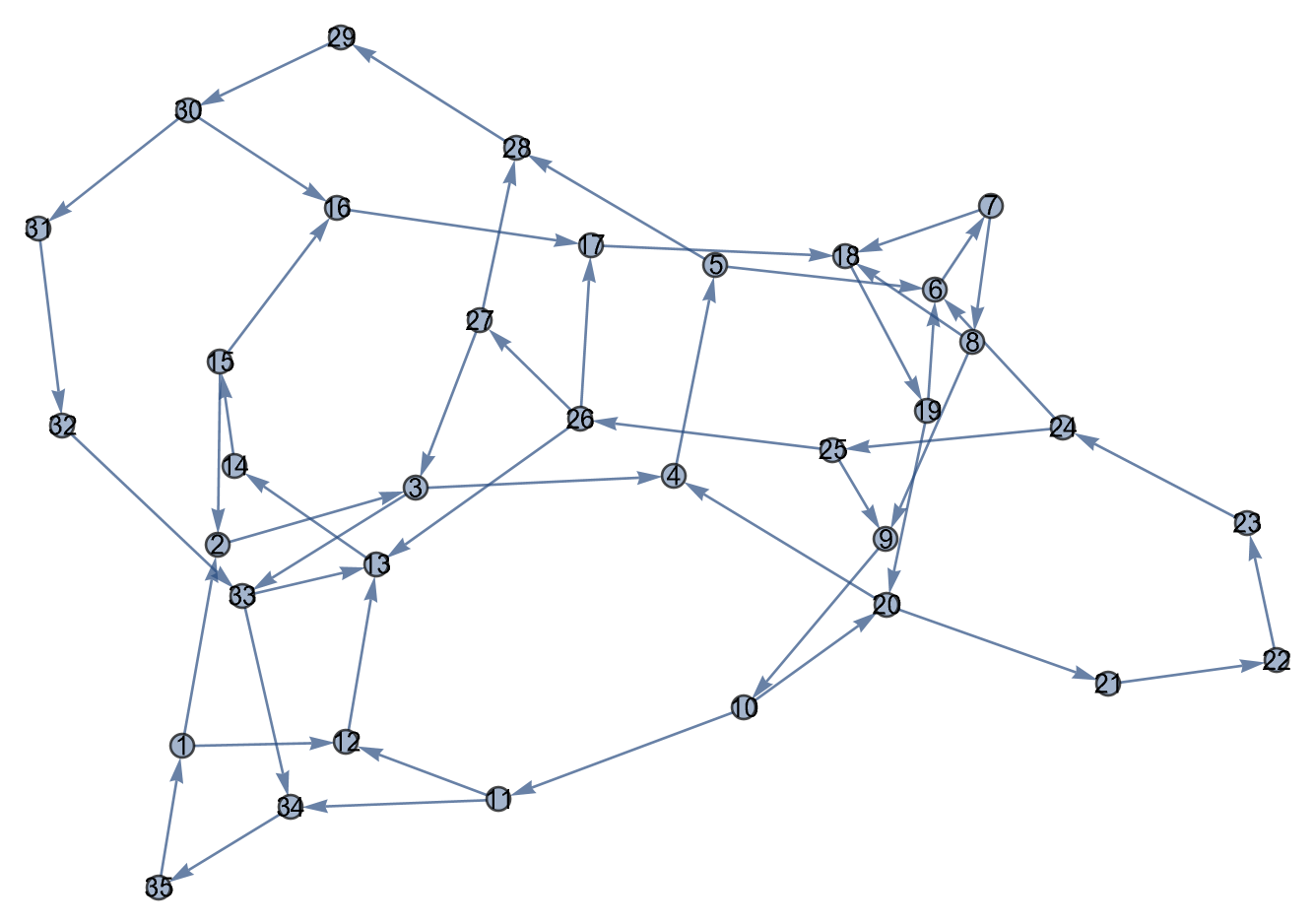}
	\caption{Directed graph with 35 vertices}
\label{fig:2023-12-21-eg351}
\end{figure}
The adjacency matrix of~$G$ and the plot of its eigenvalues are shown in figure~\ref{fig:2023-12-21-adjmat}.
 \begin{figure}[htb!]
\centering	\includegraphics[width=4.0cm, keepaspectratio]{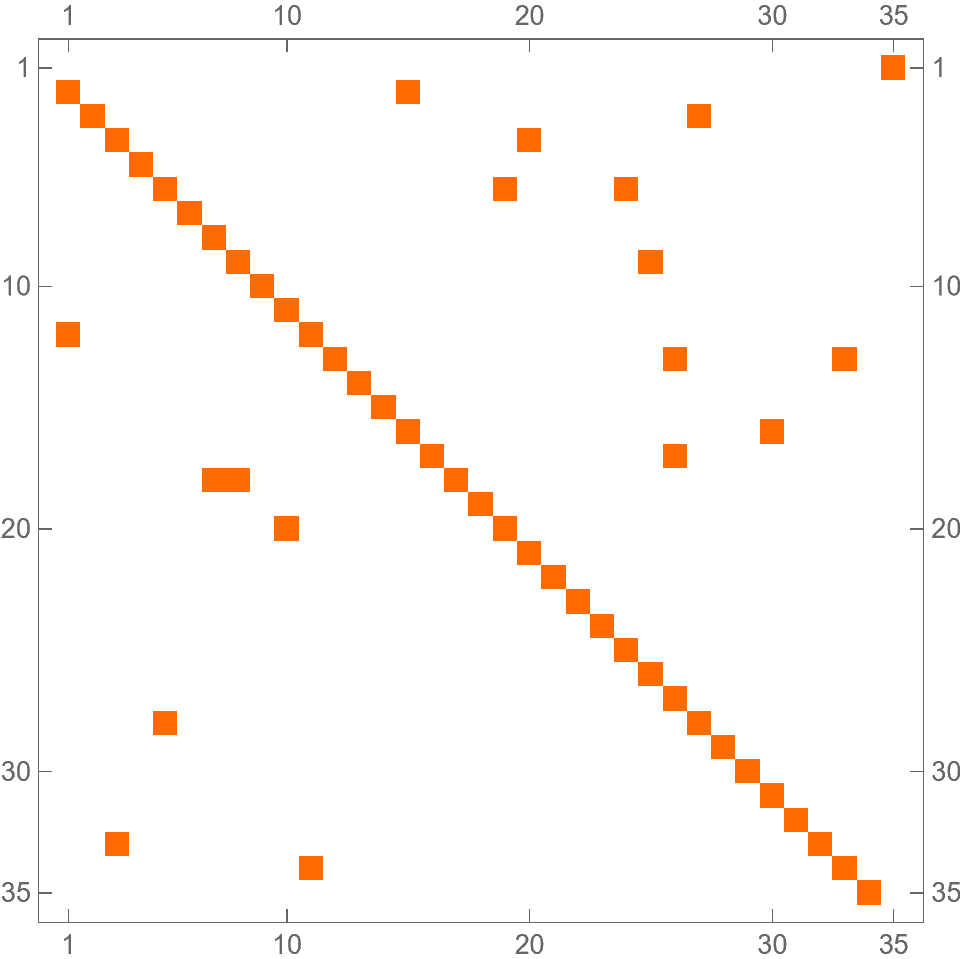}
\includegraphics[width=4.0cm, keepaspectratio]{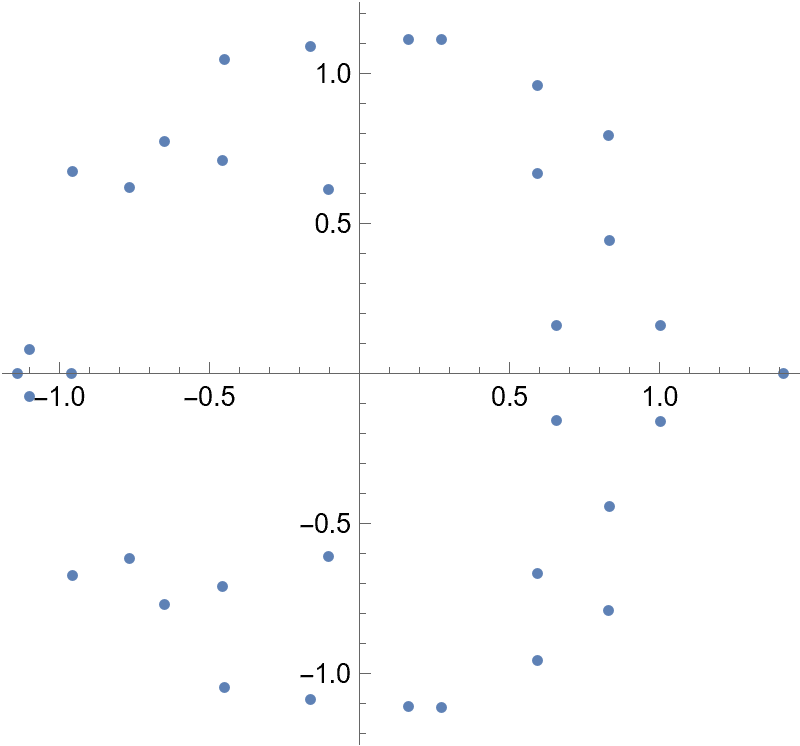}
	\caption{Graph in figure~\ref{fig:2023-12-21-eg351}: Adjacency matrix and its eigenvalues}
\label{fig:2023-12-21-adjmat}
\end{figure}

Figure~\ref{fig:2023-12-21-eg351-p0=e0-hatdelta0-delta0} illustrates for the three GSP models, at the top, the graph (vertex) signal $\mathbf{e_0}$ and, at the bottom, the graph signal $\boldsymbol{\delta}_0$ with spectrum $\frac{1}{\sqrt{N}}\mathbf{1}$. The color and dashing codes are the same for the top and bottom plots, namely, vertex, spectral, and companion representations are in dotted blue, solid green, and dashed brown:
\begin{inparaenum}[1)] \item in the top plot, the graph signal is  $\mathbf{e}_0$, i.e.,  impulsive in the vertex domain, but its $\textrm{GFT}$ is not flat, $\widehat{\mathbf{e}}_0\neq \frac{1}{\sqrt{N}}\mathbf{1}$, and its companion representation has some irregular pattern; and
\item in the bottom plot where the graph signal is an impulse in the companion domain, $\mathbf{p}=\mathbf{e}_0$, it is also flat in the spectral domain, but in the vertex domain $\boldsymbol{\delta}_0=\textrm{GFT}^{-1}\frac{1}{\sqrt{N}}\mathbf{1}$ is not impulsive.
\end{inparaenum}
 \begin{figure}[htb!]
\centering	\includegraphics[width=8.5cm, keepaspectratio]{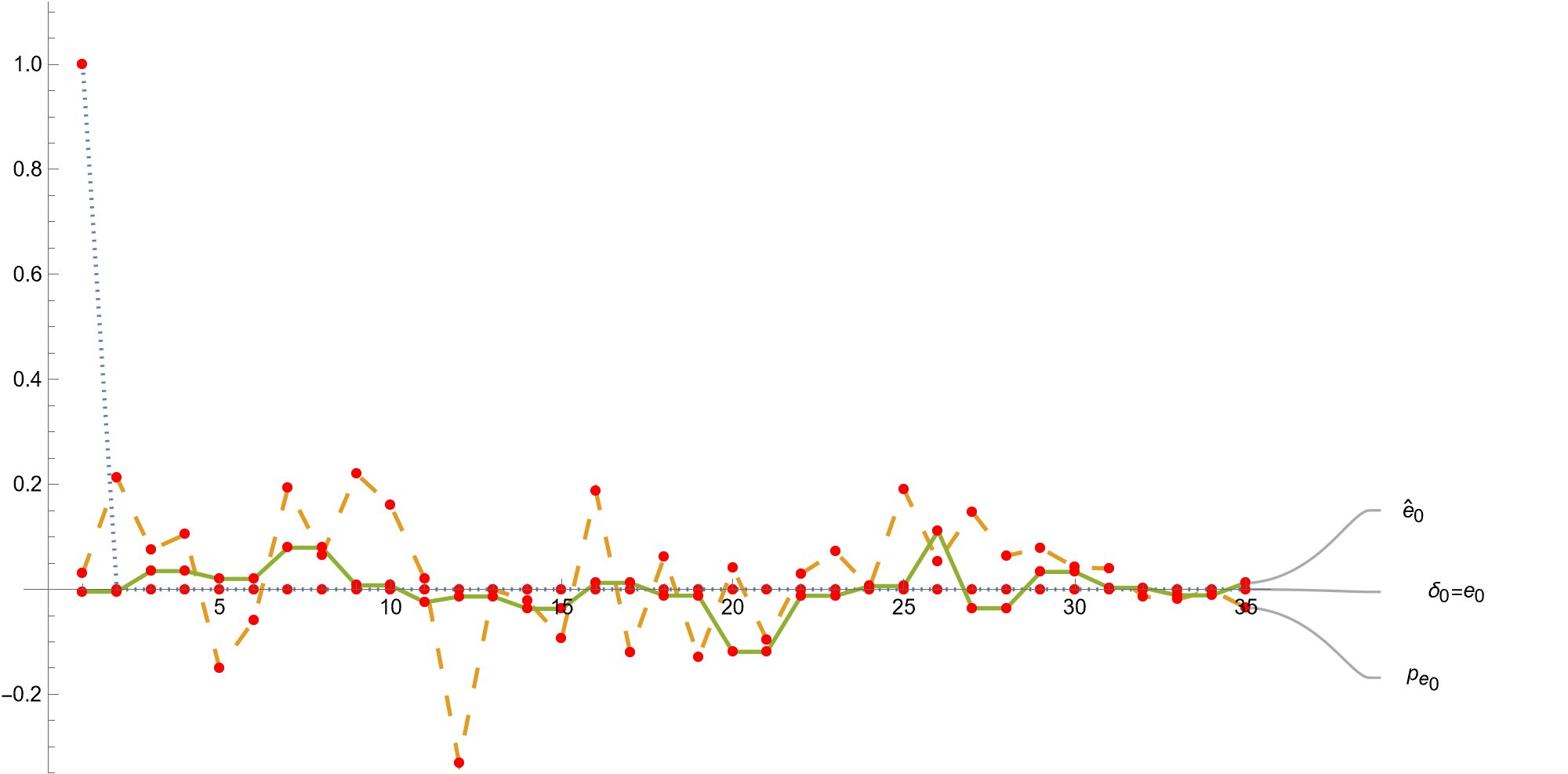}
\includegraphics[width=8.5cm, keepaspectratio]{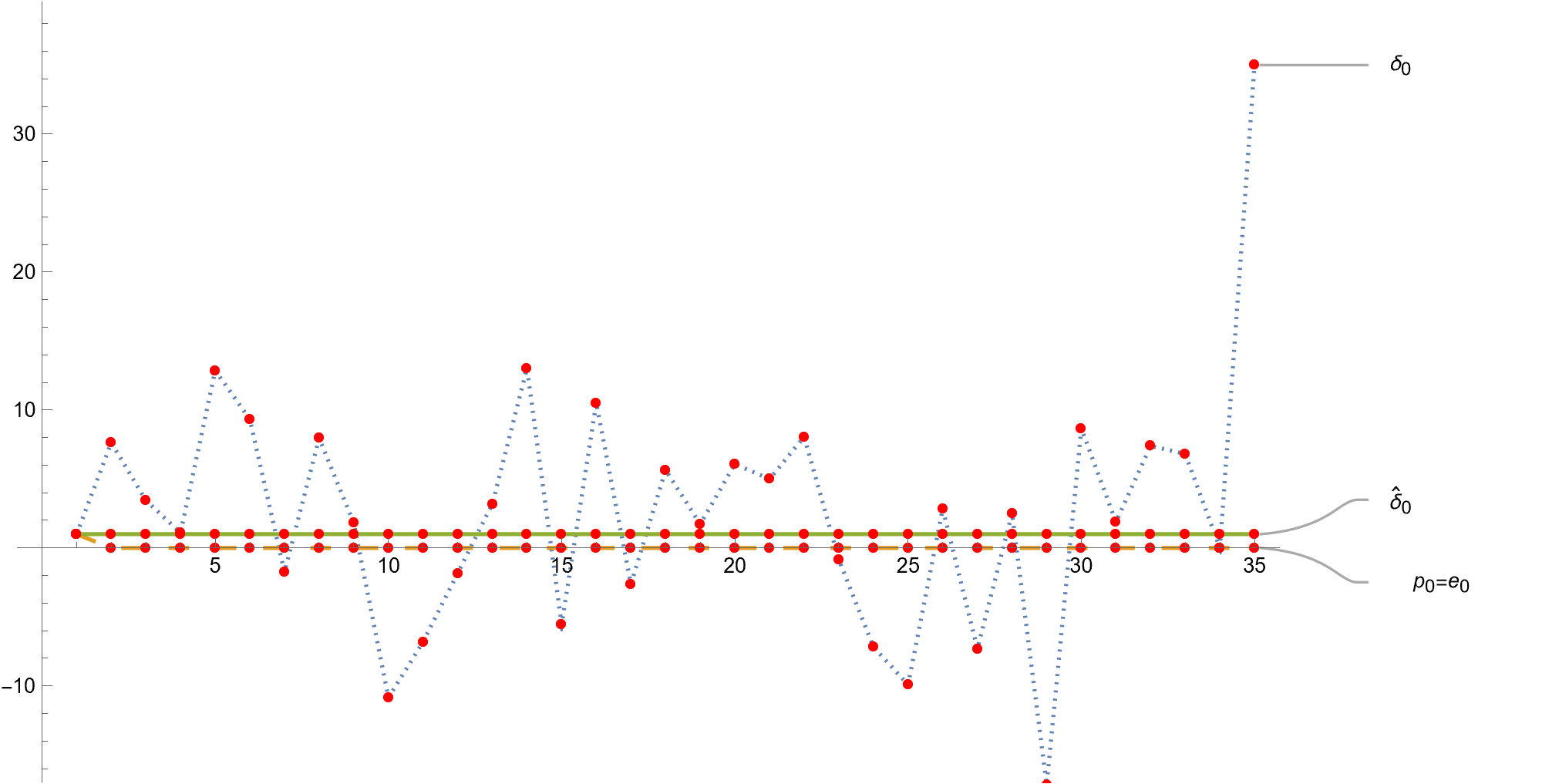}
	\caption{Dotted blue, solid green, and dashed brown lines are vertex, companion, and spectral representations. Top plot: vertex $\boldsymbol{\delta}_0=\mathbf{e}_0$, its spectrum $\widehat{\boldsymbol{\delta}}_0=\widehat{\mathbf{e}}_0$, and its companion representation  $\mathbf{p}_{\mathbf{e}_0}=\mathcal{V}^{-1}\widehat{\boldsymbol{\delta}}_0$. Bottom plot: $\widehat{\boldsymbol{\delta}}_0=\frac{1}{\sqrt{N}}\mathbf{1}$, its companion representation $\mathbf{p}_{\mathbf{e}_0}=\mathbf{e}_0$, and its vertex representation $\boldsymbol{\delta}_0$.}
\label{fig:2023-12-21-eg351-p0=e0-hatdelta0-delta0}
\end{figure}
\end{example}

\subsection{The Companion Shift}\label{subsec:companionshift}
We now consider the shift in the GSP companion model. We obtain the coordinate representation of the shift with respect to the companion basis $B_{\textrm{comp}}$ given in equation~\eqref{eqn:companionbasis} by applying the shift filter\footnote{\label{ftn:operatorandmatrixrep}We should be working here with the abstract shift~$A$ and basis vectors~$\delta_n$, but, by remarks~\ref{rmk:linearoperator-1} and~\ref{rmk:linearoperator-2}, we can work with the corresponding concepts of the coordinate matrix relative to the standard basis~$B$.}~$\mathbf{A}$ to each basis vector $\boldsymbol{\delta}_n$ and representing the result in terms of the basis vectors. By definition of the impulse and its shifted versions, see equation~\eqref{eqn:shiftedelta1-6}, we have
\begin{align}\label{eqn:companionshiftandshiftedimp-1}
\mathbf{A}\boldsymbol{\delta}_n=\mathbf{A}^{n+1}\boldsymbol{\delta}_0=\boldsymbol{\delta}_{n+1},\:\:\:\: 0\leq n\leq N-2
\end{align}
When shifting $\delta_{N-1}$, get
\begin{align}\label{eqn:companionshiftandshiftedimp-2}
A\boldsymbol{\delta}_{N-1}=A^N\boldsymbol{\delta}_0=\boldsymbol{\delta}_N.
\end{align}
The graph signal  $\boldsymbol{\delta}_N$ is not a basis vector in $B_{\textrm{comp}}$. This is the problem of \textit{signal continuation}. We address it by expressing $\boldsymbol{\delta}_N$  in terms of the basis vectors $\left\{\boldsymbol{\delta}_n\right\}_{0\leq n\leq N-1}$. Because this is a boundary problem, it is referred to as a boundary condition~(bc).

 It turns out that the bc is implicit in the choice of the shift filter.
Recall that, by the Cayley-Hamilton Theorem \cite{gantmacher1959matrix}, a linear operator defined on a vector space of dimension~$N$ satisfies its characteristic polynomial\footnote{\label{ftn:characteristicpoly-linop}The characteristic polynomial of a linear operator defined on an $N$-dimensional vector space is the monic polynomial of degree~$N$ that is zero at all eigenvalues of the operator (with multiplicity). Given remarks~\ref{rmk:linearoperator-1} and~\ref{rmk:linearoperator-2}, it is the characteristic polynomial of any matrix representing the operator relative to some basis. Since the characteristic polynomial is computed by a determinant, it is well defined because it does not depend on the basis.}
\begin{align}\label{eqn:characteristicpoly-1}
\Delta_{\mathbf{A}}(\lambda){}&=c_0+c_1\lambda+\cdots+c_{N-1}\lambda^{N-1}+\lambda^N.
\end{align}
Then
\begin{align}\label{eqn:characteristicpoly-2}
\Delta_{\mathbf{A}}(\mathbf{A}){}&=c_0\mathbf{I}+c_1\mathbf{A}+\cdots+c_{N-1}\mathbf{A}^{N-1}+\mathbf{A}^N=\mathbf{0}.
\end{align}
From~\eqref{eqn:characteristicpoly-2}, get
\begin{align}\label{eqn:characteristicpoly-3}
\mathbf{A}^N{}&=-c_0\mathbf{I}-c_1\mathbf{A}-\cdots-c_{N-1}\mathbf{A}^{N-1}.
\end{align}
Collecting the $N-1$ equations in~\eqref{eqn:companionshiftandshiftedimp-1} and equation~\eqref{eqn:characteristicpoly-3}, get the coordinate matrix of the shift for the GSP companion model:
\begin{align}\label{eqn:companionshift-1}
\mathbf{A}\cdot\left[\begin{array}{cccc}
\boldsymbol{\delta}_0\,\boldsymbol{\delta}_1\cdots \boldsymbol{\delta}_{N-1}
\end{array}\right]{}&=\\
\label{eqn:companionshift-2}
{}&\hspace{-1.9cm}\left[\!\begin{array}{cccc}
\boldsymbol{\delta}_0\,\boldsymbol{\delta}_1\cdots \boldsymbol{\delta}_{N-1}
\end{array}\!\right]\underbrace{\left[\!\begin{array}{cccc}
0&\cdots&0&-c_0\\
1&\ddots&\vdots&-c_1\\
\vdots&\ddots&0&\vdots\\
0&\cdots&1&-c_{N-1}\end{array}\!\right]}_{\mathbf{C}_{\textrm{comp}}}
\end{align}
The companion shift $\mathbf{C}_{\textrm{comp}}$  is zero everywhere, except the first lower diagonal of~1's and the last column that shows from top to bottom the negative of the coefficients from lowest to highest degree (except the coefficient of the $N$th-degree term) of the characteristic polynomial, see~\eqref{eqn:characteristicpoly-1}.

\textbf{Shifts~$\mathbf{A}$ and~$\mathbf{C}_{\textrm{comp}}$}. From equation~\eqref{eqn:companionshift-1}, we get:
\begin{align}\label{eqn:shiftingbyAandC-1}
\mathbf{A}\cdot\mathbf{s}{}&=\mathbf{A}\left[\begin{array}{cccc}
\boldsymbol{\delta}_0\,\boldsymbol{\delta}_1\cdots \boldsymbol{\delta}_{N-1}
\end{array}\right]\mathbf{p}_{\textrm{comp}}\\
\label{eqn:shiftingbyAandC-1}
{}&=\phantom{\mathbf{A}}\left[\begin{array}{cccc}
\boldsymbol{\delta}_0\,\boldsymbol{\delta}_1\cdots \boldsymbol{\delta}_{N-1}
\end{array}\right]\mathbf{C}_{\textrm{comp}}\,\mathbf{p}_{\textrm{comp}}
\end{align}
This shows that shifting signal~$\mathbf{s}$ by~$\mathbf{A}$ shifts~$\mathbf{p}_{\textrm{comp}}$ by~$\mathbf{C}_{\textrm{comp}}$.

\textbf{Structure of the GSP companion model shift }. From~\eqref{eqn:companionshift-2}, decompose $\mathbf{C}_{\textrm{comp}}$ as
\begin{align}\label{eqn:structureofCcomp-1}
\mathbf{C}_{\textrm{comp}} {}&= \left[\begin{array}{cccc}
0&\cdots&0&-c_0\\
1&\ddots&\vdots&-c_1\\
\vdots&\ddots&0&\vdots\\
0&\cdots&1&-c_{N-1}\end{array}\right]\\
\nonumber\\
\label{eqn:structureofCcomp-2}
{}&=\underbrace{\left[\!\begin{array}{cccc}
0&\cdots&0&0\\
1&\ddots&\vdots&0\\
\vdots&\ddots&\ddots&\vdots\\
0&\cdots&1&0\end{array}\!\right]}_{\mathbf{A}_{\textrm{path}}}\!+\!\underbrace{\left[\!\begin{array}{cccc}
0&\cdots&0&\hspace{-.5cm}-c_0\\
0&\ddots&\vdots&\hspace{-.5cm}-c_1\\
\vdots&\ddots&\ddots&\hspace{-.5cm}\vdots\\
0&\cdots&0&\hspace{-.2cm}-c_{N-1}\end{array}\!\right]}_{\mathbf{C}_{\textrm{bc}}}
\end{align}
The first term $\mathbf{A}_{\textrm{path}}$ is the shift associated with a $N$th-node directed path, and the second term $\mathbf{C}_{\textrm{bc}}$ is the bc term of the companion shift $\mathbf{C}_{\textrm{bc}}$.

In alternative, we can rewrite~\eqref{eqn:structureofCcomp-1} as
\begin{align}\label{eqn:structureofCcomp-3}
\mathbf{C}_{\textrm{comp}} {}&\!= \!\underbrace{\left[\!\begin{array}{cccc}
0&\cdots&0&1\\
1&\ddots&\vdots&0\\
\vdots&\ddots&\ddots&\vdots\\
0&\cdots&1&0\end{array}\!\right]}_{\mathbf{A}_{\textrm{cyclic}}}\!+\!\underbrace{\left[\!\begin{array}{cccc}
0&\cdots&0&\hspace{-.5cm}-c_0-1\\
0&\ddots&\vdots&\hspace{-.5cm}-c_1\\
\vdots&\ddots&\ddots&\hspace{-.5cm}\vdots\\
0&\cdots&0&\hspace{-.2cm}-c_{N-1}\end{array}\!\right]}_{\mathbf{C}^\prime_{\textrm{bc}}}\\
\label{eqn:structureofCcomp-4}
{}&=\mathbf{A}_{\textrm{cyclic}}+\left[\begin{array}{c}
-c_0-1\\
-c_1\\
\vdots\\
-c_{N-1}\end{array}\right]\mathbf{e}_{N-1}^T
\end{align}
Equations~\eqref{eqn:structureofCcomp-3} and~\eqref{eqn:structureofCcomp-4} show that the companion shift $\mathbf{C}_{\textrm{comp}}$ equals the cyclic shift $\mathbf{A}_{\textrm{cyclic}}$ of the directed cycle shift \cite{Sandryhaila:13} associated with (periodic) time signals plus a bc $\mathbf{C}^\prime_{\textrm{bc}}$. This confirms that GSP can be interpreted like DSP plus an appropriate boundary condition.

\subsection{Companion graph}\label{subsec:companiongraph}
\begin{figure}[htb!]
\centering	\includegraphics[width=8.5cm, keepaspectratio]{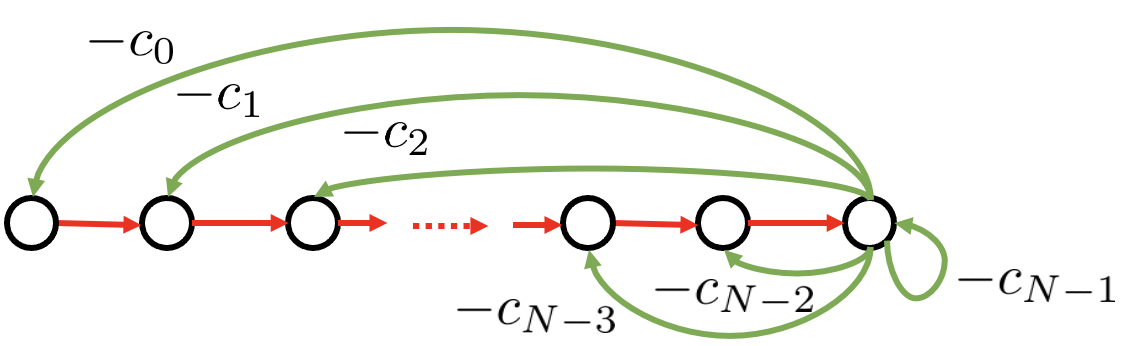}
\includegraphics[width=8.5cm,height=4cm]{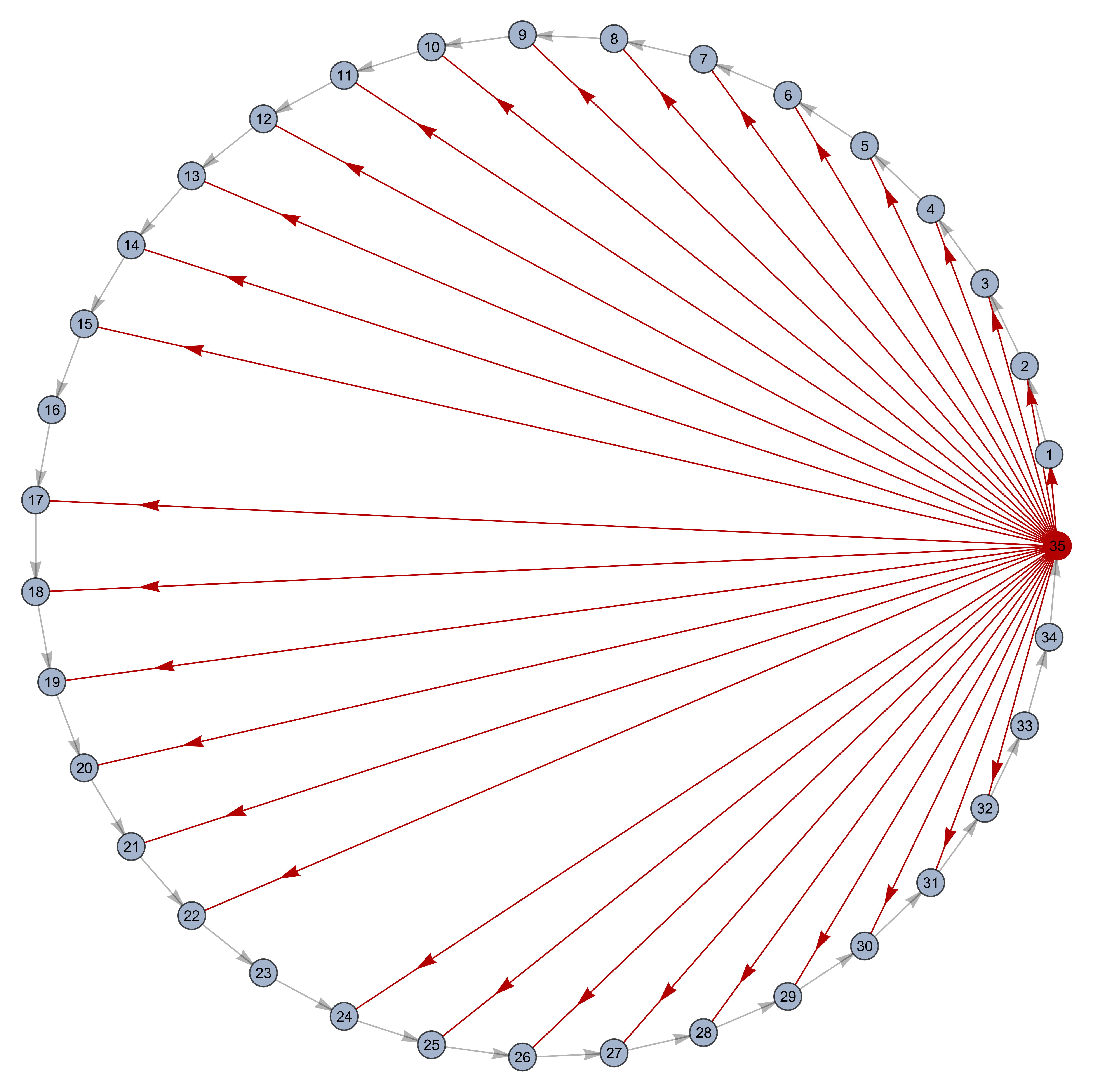}
	\caption{Companion graph. Generic graph at the top:  unlabelled edges in red of the directed path have weight~1. Edges in green reflect the bc and are  labeled by their weights. Companion graph for \textit{directed} graph in example~\ref{exp:impulsivesignals} at the bottom: Nodes are labeled from~1 to~35. Node~35 and weighted edges from this to other nodes in red. Their weights, not shown, are the coefficients of $\Delta_{\mathbf{A}}(x)$ given in example~\ref{exp:impulsivesignals}.}
		\label{fig:Sgraph}
\end{figure}

As noted before, in GSP the adjacency matrix of a graph~$G$ defines the GSP shift. Likewise, the companion shift  $\mathbf{C}_{\textrm{comp}}$ can be considered as the weighted adjacency matrix of the graph in the GSP companion model.  The companion graph $G_{\textrm{comp}}=\left(V_{\textrm{comp}}, E_{\textrm{comp}}\right)$ for a generic adjacency graph is displayed on top of figure~\ref{fig:Sgraph} and for the graph in example~\ref{exp:impulsivesignals} at the bottom of the figure.
The following holds:
\begin{inparaenum}[1)]
\item\label{inp:Gcom-1} the node set $V_{\textrm{comp}}$ of the \textit{companion} graph $G_{\textrm{comp}}$ has~$N$ nodes like the node set~$V$ of the original graph~$G$;
\item  \label{inp:Gcom-1-1} the nodes of $V_{\textrm{comp}}$ are different from the nodes of~$V$ of~$G$;
\item \label{inp:Gcom-2} each of the $N$~nodes of $G_{\textrm{comp}}$ corresponds to a basis vector of $B_{\textrm{comp}}$, i.e., to one of the shifted $\boldsymbol{\delta}_n$ just like each node in~$V$ corresponds to a standard vector $\mathbf{e}_n$;
\item \label{inp:Gcom-3} the companion graph $G_{\textrm{comp}}$ is directed and weighted;
\item \label{inp:Gcom-4} the structure of $G_{\textrm{comp}}$ echoes  the structure of the companion shift $\mathbf{C}_{\textrm{comp}}$ given in equation~\eqref{eqn:structureofCcomp-2}:
\begin{inparaenum}[\ref{inp:Gcom-4}.i)]
\item \label{inp:Gcom-4.i} A directed path of~$N$ nodes with adjacency matrix $\mathbf{C}_{\textrm{path}}$, and
\item \label{inp:Gcom-4.ii} reflecting the bc, a set of weighted directed edges from the most right node $N-1$ to the other nodes, possibly including a self-loop if $c_{N-1}\neq 0$;
\end{inparaenum}
\item \label{inp:Gcom-5} iff $c_0\neq 0$, the companion graph is strongly connected. This is the case if zero is not an eigenvalue of~$\mathbf{A}$.
\end{inparaenum}

Under assumption~\ref{ass:Adistincteigenvalues}, \textit{any} directed or undirected graph~$G$ has a corresponding weighted \textit{companion} graph. Under assumption~\ref{ass:Adistincteigenvalues}, they are unique. Hence, both, $G_{\textrm{comp}}$ and $\mathbf{C}_{\textrm{comp}}$, are \textit{canonical} representations connected with \textit{any} GSP model satisfying assumption~\ref{ass:Adistincteigenvalues}.
\begin{example}[Time graph: Directed cycle graph]\label{exp:cycliggraph}
 The structure of the companion graph in figure~\ref{fig:Sgraph} extends the structure of the DSP directed cyclic graph \cite{Sandryhaila:13} whose adjacency matrix is $\mathbf{A}_{\textrm{cyclic}}$ given in equation~\eqref{eqn:structureofCcomp-3}. The DSP cyclic graph  follows from the companion graph at the top of  figure~\ref{fig:Sgraph} by taking $c_0=-1$ and eliminating the self-loop and all the remaining backward pointing edges.
\end{example}

\begin{example}[DCT-II graph]\label{exp:DCT-IIgraph}
While example~\ref{exp:impulsivesignals} worked with a directed graph, this example shows that the theory applies equally to undirected graphs. While we work this example with the adjacency matrix, we could similarly work with the graph Laplacian. Figure~\ref{fig:dctII} shows on the left the undirected graph whose graph Fourier transform is the (1D)-DCT-II \cite{Pueschel:08b} and on the right its companion graph. We see the unweighted directed path with 35 nodes and, like with the graph at the bottom of  figure~\ref{fig:Sgraph}, node 35 and the weighted edges from node 35 to the other nodes are in red. The weights of these edges are the coefficients of the characteristic polynomial
$\Delta_{\mathbf{A}_{\textrm{DCT-II}}}(x)=1. x^{35}-1. x^{34}-8.25 x^{33}+8.25 x^{32}+31. x^{31}-31. x^{30}-70.23 x^{29}+70.23 x^{28}+107.05 x^{27}-107.05 x^{26}-115.97 x^{25}+115.97 x^{24}+91.98 x^{23}-91.98 x^{22}-54.20 x^{21}+54.20 x^{20}+23.84 x^{19}-23.84 x^{18}-7.79 x^{17}+7.79 x^{16}+1.87 x^{15}-1.87 x^{14}-0.32 x^{13}+0.32 x^{12}+0.039 x^{11}-0.039 x^{10}-0.003 x^9+0.003 x^8+0.0001 x^7-0.0001 x^6-3.61\times 10^{-6}x^5+3.61\times 10^{-6} x^4+3.56 x^3\times 10^{-8}-3.56 x^2\times 10^{-8}-5.82\times 10^{-11} x+5.82\times 10^{-11}$.
\begin{figure}[htb!]
\centering	\includegraphics[width=4.0cm, keepaspectratio]{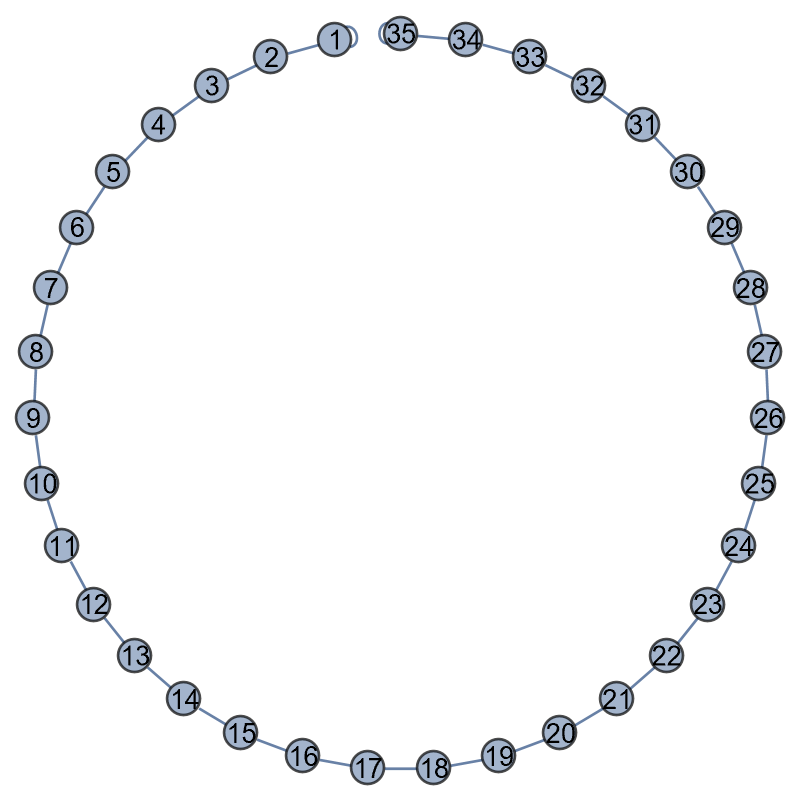}
\centering	\includegraphics[width=4.5cm, keepaspectratio]{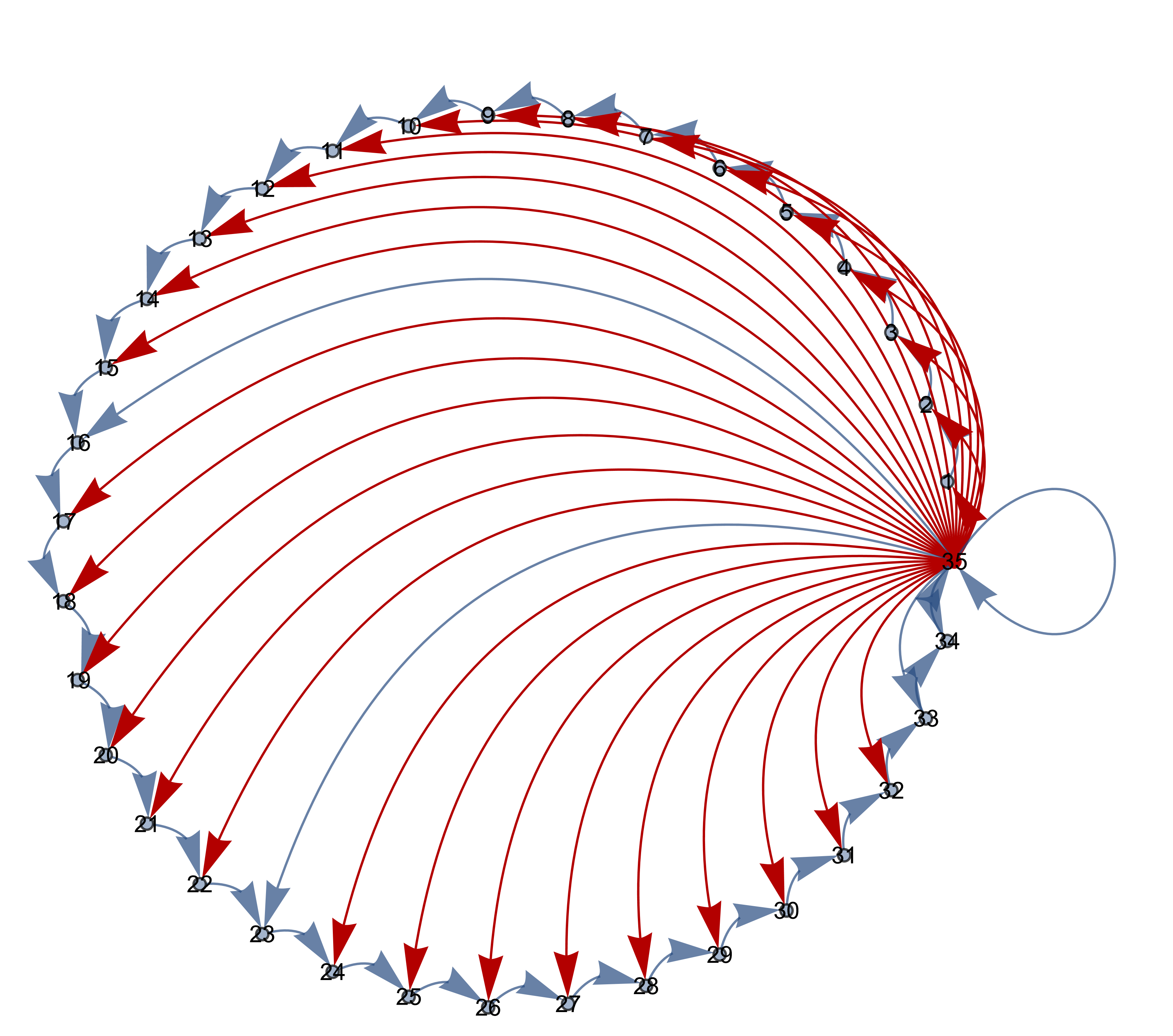}	\caption{DCT-II (\textit{undirected}) graph on 35 nodes and its companion graph.}	\label{fig:dctII}
\end{figure}
\end{example}

\begin{example}[Ladder graph]
Figure~\ref{fig:ladderdir12} shows on top a ``directed'' ladder graph with 12 nodes and below it the corresponding companion graph. The characteristic polynomial of the adjacency matrix of a ladder graph like the one shown in the figure with $N=2K$ nodes is
\begin{align}\label{eqn:Deltaladderdirected12}
\Delta_A(x){}&=-1-x^2-x^4-x^8-\cdots-x^{2(K-2)}+x^{2K}.
\end{align}
The polynomial $\Delta_A(x)$ shows that the nonzero edge weights of the companion graph of the directed ladder graph are all ones (the nonzero coefficients of $\Delta_A(x)$ are  $c_{n}\equiv-1$, $n\neq 2K$).
\begin{figure}[htb!]
\centering	\includegraphics[width=6cm, keepaspectratio]{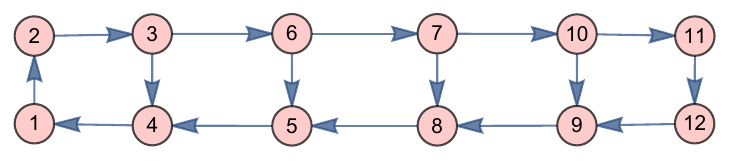}\\
\centering	\includegraphics[width=6cm, keepaspectratio]{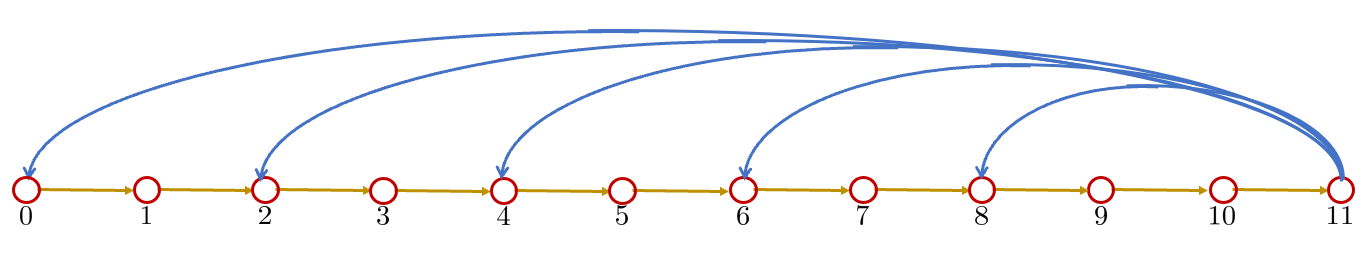}	\caption{Directed ladder graph and its companion graph.}	\label{fig:ladderdir12}
\end{figure}
\end{example}

These examples illustrate that the companion graph, regardless if for directed or undirected graph signal models, exhibits the structure of a directed weighted graph with a path of unweighted edges and a number of bc weighted edges from the last node to previous nodes where the weights are obtained from the coefficients of the characteristic polynomial $\Delta_{\mathbf{A}}(x)$ of the original model graph.
%

\subsection{Graph Fourier Transform in the Companion Model}\label{subsec:GFTinCompModel}
Before leaving the section, we consider the graph Fourier transform in the GSP companion model. We start by diagonalizing the companion shift. It is well known that it is diagonalized by the Vandermonde matrix~$\mathcal{V}$
\begin{align}\label{eqn:diagonalizationcompshift-1}
\mathbf{C}_{\textrm{comp}}{}&=\mathcal{V}^{-1}\boldsymbol{\Lambda}\mathcal{V},
\end{align}
where as before $\mathcal{V}$ is the Vandermonde matrix and~$\boldsymbol{\Lambda}$ is diagonal matrix of the eigenvalues. The $\textrm{GFT}$ of the companion model is the Vandermonde matrix~$\frac{1}{\sqrt{N}}\mathcal{V}$ and, by~\eqref{eqn:vandermonde-6}, since $\frac{1}{\sqrt{N}}\mathcal{V}\mathbf{p}_{\textrm{comp}}=\widehat{\mathbf{s}}$, the graph spectral domain in the companion modal is actually the (vertex) graph spectral domain.
\section{Vertex, Spectral, and Companion GSP Models}\label{sec:vertexspcompanion}
Table~\ref{tab:vertexspcompanion} summarizes the main ingredients of the three GSP models that we have considered: the coordinate vectors with respect to the basis vectors, the diagonalized shifts, the polynomial filters, the graph underlying the models, the graph Fourier transforms, and the confusion factors that relate the companion model quantities like the signal representation~$\mathbf{p}_{\textrm{comp}}$ to the standard and spectral model quantities like~$\mathbf{s}$ and~$\widehat{\mathbf{s}}$.
\begin{table}
    \centering
{\scriptsize    \begin{tabular}{|P{1.7cm}|P{1.8cm}|P{1.6cm}|P{1.95cm}|}
    \hline
&Vertex&Spectral&Companion  \\\hline
Signal&$\mathbf{s}$&$\widehat{\mathbf{s}}$&$\mathbf{p_{\textrm{comp}}}$\\\hline
Basis&$\left\{\mathbf{e}_n\right\}$&$\left\{\mathbf{v}_n\right\}$&$\left\{\boldsymbol{\delta}_n\right\}$\\\hline
\textrm{Shift}&$\mathbf{A}=\mathbf{V}\boldsymbol{\Lambda}\mathbf{V}^{-1}$&$\mathbf{M}=\mathbf{V}^{-1}\boldsymbol{\Lambda}^*\mathbf{V}$&$\mathbf{C}_{\textrm{comp}}=\mathcal{V}^{-1}\boldsymbol{\Lambda}\mathcal{V}$\\\hline
LSI filters&$P(\mathbf{A})$&$P(\mathbf{M})$&$P(\mathbf{C}_{\textrm{comp}})$
\\\hline
Graph&$G$&$G_{\textrm{sp}}$&$G_{\textrm{comp}}$\\\hline
GFT&$\mathbf{V}^{-1}$&$\mathbf{V}$&$\frac{1}{\sqrt{N}}\mathcal{V}$
\\\hline
Confusion factor&$\mathbf{p}_{\textrm{comp}}\xrightarrow{\frac{1}{\sqrt{N}}\mathbf{V}\mathcal{V}}\mathbf{s}$&$\mathbf{p}_{\textrm{comp}}\xrightarrow{\frac{1}{\sqrt{N}}\mathcal{V}}\widehat{\mathbf{s}}$&\\\hline
    \end{tabular}
    \caption{Vertex, spectral, companion GSP models}
\label{tab:vertexspcompanion}
    }
\end{table}
We comment on the nature of these models.
\begin{enumerate}
\item The vertex model is the natural one, where the signal samples are commonly indexed by the agent~$n$ that measures, observes, or collects them\textemdash a sensor, an individual, a stock, a commodity\textemdash and the signal is the sum of these $N-1$ observations (samples) $\mathbf{s}=\sum s_n \mathbf{e}_n$.
\item The spectral model decomposes the signal space, the module~$\mathcal{M}$, into irreducible invariant signal modules (subspaces) that, under assumption~\ref{ass:diagonalizableA}, are one dimensional (1D) and are the eigenspaces of the LSI filters ($h(v)=h(\lambda)\cdot v$). The GFT is the transform that maps~$\mathcal{M}$ onto the direct sum of its irreducible invariant subspaces (eigenspaces).
\item Under assumption~\ref{ass:Adistincteigenvalues}, $m_\mathbf{A}(x)=\Delta_\mathbf{A}(x)$, i.e., the minimal polynomial $m_\mathbf{A}(x)$ \cite{gantmacher1959matrix} of the shift~$\mathbf{A}$ equals its characteristic polynomial $\Delta_\mathbf{A}(x)$. Then, the GSP signal space module~$\mathcal{M}$ is cyclic, i.e., it is generated by a single vector and its $N-1$ shifts\textemdash for the companion model, the signal space module is $\mathcal{M}=\textrm{span}\!\!-\!\!\left\{\boldsymbol{\delta}_0,\mathbf{A}\cdot \boldsymbol{\delta}_0, \cdots,\mathbf{A}^{N-1}\boldsymbol{\delta}_0\right\}$. The companion model recovers for GSP the cyclic nature of DSP.
\end{enumerate}
Each of these descriptions of the graph signal has their own advantages: the vertex model is the natural way by which signals are commonly obtained; the spectral model decomposes signals into their 1D eigencomponents, invariant to LSI filters; and the companion GSP model parallels DSP the closest, enabling  successful concepts and algorithms from DSP to be ported to GSP through the S$^2$CM$^2$ method.

In contrast to GSP, in DSP, the vertex and companion models are the same, $\mathbf{s}=\mathbf{p}_{\textrm{comp}}$, $B=B_{\textrm{comp}}$, $\mathbf{e}_n=\boldsymbol{\delta}_n$, $\mathbf{A}=\mathbf{C}_{\textrm{comp}}$, and the $\textrm{DFT}=\frac{1}{\sqrt{N}}\mathcal{V}$ is the Vandermonde matrix, so that the confusion factors from the companion domain to the vertex and spectral domains are the identity and the \textrm{DFT}, respectively.
\section{The Companion Model and the Step~1-Step~2 Companion Model Method (S$^2$CM$^2$)}\label{sec:compmodS2CM2}
As observed, the companion model exploits the \textit{cyclic} nature of the signal module~$\mathcal{M}$ under assumption~\ref{ass:Adistincteigenvalues}. This parallels the cyclic nature of DSP for which $\boldsymbol{\delta}_0=\mathbf{e}_0$ and $B_{\textrm{comp}}=B$, i.e., the companion and standard basis~$B$ coincide. This shows that the GSP companion model replicates the cyclic nature of the time DSP model and enables extending effortless GSP with new concepts, algorithms, and methods as we now illustrate with a few examples. We call this the Step~1-Step~2 Companion Model Method (S$^2$CM$^2$) where Step~1 introduces the new GSP concept by adopting its DSP version in the companion model, and Step~2 maps the companion concept to the vertex domain using the \textit{confusion} factor $\frac{1}{\sqrt{N}}\textrm{GFT}^{-1}\mathcal{V}=\frac{1}{\sqrt{N}}\mathbf{V}\mathcal{V}$ as shown in figure~\ref{fig:3signaldomains-1} or table~\ref{tab:vertexspcompanion}.
\subsection{Impulse}\label{subsec:impulse}
In previous sections, we have introduced the graph impulse as the inverse GFT of a graph signal with flat spectrum. This seemed to be an arbitrary choice, and one may ask why not choose as graph impulse the graph signal $\mathbf{e}_0$ that is impulsive in the vertex domain. But with this choice,   the shifts $\mathbf{A}^n\mathbf{e}_0$ are not impulsive, and the spectrum $\widehat{e}_0$ is not flat.

 If we apply the S$^2$CM$^2$ method, we choose as impulse the graph signal in the companion domain defined by $\boldsymbol{\delta}_0=\mathbf{p}_0=\mathbf{e}_0$. In the companion model:
\begin{inparaenum}[1)]
\item \label{inp:delta0-e0} this signal is impulsive, see also for example   figure~\ref{fig:2023-12-21-eg351-p0=e0-hatdelta0-delta0};
\item  \label{inp:hatdelta0-flat}  its spectrum is flat $\widehat{\boldsymbol{\delta}}_0=\frac{1}{\sqrt{N}}\mathcal{V}\cdot \mathbf{e}_0=\frac{1}{\sqrt{N}}\mathbf{1}$, since the first column of $\mathcal{V}$ is the vector $\boldsymbol{\lambda}^0=\mathbf{1}$;
 \item  \label{inp:delta0n-impulsive} we can verify by direct computation that its shifts $\boldsymbol{\delta}_n=\mathbf{C_{\textrm{comp}}}^n\boldsymbol{\delta}_0=\mathbf{e}_n$, $0\leq n\leq N-2$, are impulsive, and
\item  \label{inp:hatdelta0n-lambdan}
 the spectrum of its shifts are ``phase changes,'' $\widehat{\boldsymbol{\delta}}_n=\boldsymbol{\lambda}^n$, of the flat spectrum $\widehat{\delta}_0=\frac{1}{\sqrt{N}}\mathbf{1}$.
\end{inparaenum}
By applying Step~1 of S$^2$CM$^2$, we see that the choice $\mathbf{p}_0=\mathbf{e}_0$ does replicate all the DSP characteristics of an impulse. To obtain the vertex representation of the impulse, by  Step~2, we apply the confusion factor and get $\frac{1}{\sqrt{N}}\mathbf{V}\mathcal{V}\mathbf{e}_0=\mathbf{V}\frac{1}{\sqrt{N}}\mathbf{1}$. This gives comfort to   our previous choice of the graph impulse.

 \begin{remark}[Cyclic generator for signal space~$\mathcal{M}$] \label{rmk:delta0cyclicgenerator} Under  assumption~\ref{ass:Adistincteigenvalues} of distinct eigenvalues, the set $B_{\textrm{comp}}$ of the impulse $\boldsymbol{\delta}_0$ and its shifts is the companion basis. This shows that $\boldsymbol{\delta}_0$ is a cyclic generator of the signal module~$\mathcal{M}$. Since it corresponds to the graph signal with flat  spectrum,
items~\ref{inp:hatdelta0-flat} and~\ref{inp:delta0n-impulsive} confirm that for diagonalizable shift with distinct eigenvalues the sum of the eigenvectors is a cyclic generator of the signal space~$\mathcal{M}$ \cite{hoffmankunze}.
\end{remark}
\subsection{GSP $z$-Transform}\label{subsec:GSPz-transform}
We are not aware of a successful definition of the GSP $z$-transform of a graph signal. Attempting it  directly in the vertex domain is not promising. Instead, we resort to S$^2$CM$^2$.

By Step~1, we first rewrite the companion representation of graph signals as polynomials. Let~$\mathbf{p}_s$ be the companion representation for signal~$\mathbf{s}$. Writing it as a polynomial
\begin{align}\label{eqn:polyrep-1}
p_s(x){}&=p_0+p_1x+\cdots+p_{N-1}x^{N-1},
\end{align}
where the coefficients of the polynomial $p_s(x)$ are the entries (or companion ``samples'') of the companion representation $\mathbf{p}_s$, and the  indeterminate~$x$ represents the basis vectors of $B_{\textrm{comp}}$, i.e., the impulse and its shifts. The polynomial $p_s(x)$ is of degree at most $N-1$.

We now interpret~$x$ as $z^{-1}$. By Step~1, $p_s(x)$ is then the GSP $z$-transform of the graph signal~$\mathbf{s}$. Note that, to get the $z$-transform, we need to go from the vertex to the companion representation $\mathbf{s}\xrightarrow{\sqrt{N}\mathcal{V}^{-1}\mathbf{V}^{-1}}\mathbf{p}_s$. It is in the companion model that the coefficients of $p_s(x)$, and so of the $z$-transform, are the ``samples'' of the signal in~\eqref{eqn:polyrep-1}. In DSP, the vertex and companion models coincide, and so the coefficients of the DSP $z$-transform are the vertex samples of the signal.

By Step~2, the $z$-transform in the vertex domain is obtained by transforming back the companion impulses to their vertex representation through the confusion factor $\frac{1}{\sqrt{N}}\mathbf{V}\mathcal{V}$.

\subsection{FFT and Fast Convolution in the GSP Companion Model}\label{subsec:fastconvolution}
This section applies S$^2$CM$^2$ to obtain fast convolution in the companion model through application of the fast Fourier transform (FFT).

Consider two graph signals~$\mathbf{s}$ and~$\mathbf{t}$ with companion representations $\mathbf{p}_s$ and $\mathbf{p}_t$ with GSP $z$-transforms, $p_s(x)$ and $p_t(x)$ as given by~\eqref{eqn:polyrep-1}.
%
\subsubsection{Companion model: Fast \textit{linear} convolution}\label{subsubsec:fastlinear convolution} By Step~1, and recalling that DSP linear convolution is obtained by the product of the $z$-transforms, the GSP linear convolution of the two signals~$\mathbf{s}$ and~$\mathbf{t}$ in the GSP companion domain is
\begin{align}\label{eqn:GSPlinearconvolution-1}
   p_u(x){}&= \left(p_t\ast p_s(x)\right)=p_t(x)\cdot p_s(x)
\end{align}
where $p_u(x)$ is the GSP $z$-transform of the linear convolution in the companion domain of the two signals~$\mathbf{s}$ and~$\mathbf{t}$. Note that the product of polynomials being commutative, the convolution given by~\eqref{eqn:GSPlinearconvolution-1} is commutative.

\textbf{Companion model: Fast linear convolution with the FFT}. Just like with DSP, GSP linear convolution, being the product of two polynomials, can be achieved by (conventional) FFT, an $(N\log N)$ operation.

To get convolution in the vertex domain, we apply Step~2, multiplication by $\frac{1}{\sqrt{N}}\mathbf{V}\mathcal{V}$, which of course is not a fast operation.

\subsubsection{Companion model: Fast \textit{circular} convolution}\label{subsubsec:fastmodular convolution}
The linear convolution~\eqref{eqn:GSPlinearconvolution-1} results in general in polynomials $p_u(x)=(p_t\ast p_s)(x)$ of degree greater than $N-1$. By Step~1, modular convolution maps it back to a polynomial of degree at most $N-1$. This is GSP circular convolution in the companion model
\begin{align}
\label{eqn:GSPmodularconvolution-1}
(p_t\oast p_s)(x){}&= \left(p_t(x)\cdot p_s(x)\right)\!\!\!\!\!\mod\! \Delta_{\textbf{A}}(x)
\end{align}
that is achieved by modularizing the result with the characteristic polynomial $\Delta_{\textbf{A}}(x)$ that captures the bc. Again, by Step~2, and application of the confusion factor, we get circular convolution in the vertex domain.

\begin{figure}[htb!]
\centering	
\includegraphics[width=8.5cm, keepaspectratio]{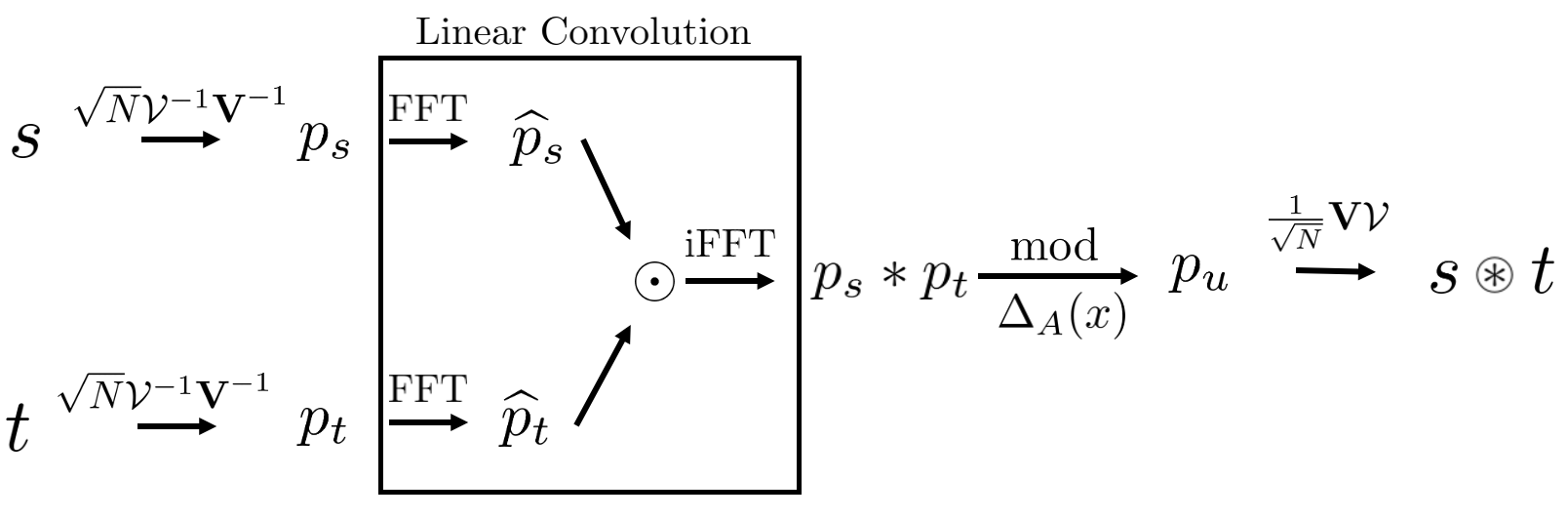}
	\caption{GSP circular convolution $s$\textcircled{$*$}$t$ by FFT.}
\label{fig:fftexample}
\end{figure}

\textbf{Fast companion model convolution by FFT}. 
  The  product of $p_s(x)$ and $p_t(x)$ can be computed by fast Fourier transform (FFT) and  the \hspace{-.185cm}$\mod$ \hspace{-.175cm} reduction by (fast) polynomial division ($O(N)$ operation \cite{Knuth81}). So, in the companion model, convolution is $O(N\ln N)$. Equations~\eqref{eqn:GSPlinearconvolution-1} and~\eqref{eqn:GSPmodularconvolution-1} are pleasing. They evaluate circular convolution in the companion model by linear convolution through the FFT, an intrinsically DSP algorithm, and then a fast $O(N)$ polynomial division algorithm \cite{Knuth81}. Figure~\ref{fig:fftexample} illustrates vertex convolution through the companion model.

We emphasize that convolution in the vertex domain is not $O(N\log N)$ nor fast because of the confusion factors between domains and the companion domain.
%
\subsection{Modulation}
\label{subsec:modulation}
This section applies the S$^2$CM$^2$ to develop GSP modulation and GSP frequency division multiplexing. DSP modulation \cite{oppenheimwillsky-1983,siebert-1986,oppenheimschaffer-1989,mitra-1998} is multiplication of a signal~$\mathbf{s}$ by a carrier signal in the time domain. Its effect (such as in frequency multiplexing) is translation or shifting in the frequency domain.

The DSP carriers are the harmonic \textit{eigenvectors} $\mathbf{v}_n$ of the time shift, i.e., the eigenvectors of the cyclic directed shift $\mathbf{A}_{\textrm{cyclic}}$, see equation~\eqref{eqn:structureofCcomp-3}. If we were to extrapolate this to the GSP vertex model, we would choose as carriers in GSP modulation the eigenvectors $\mathbf{v}_n$ of the GSP vertex shift~$\mathbf{A}$. This is done, for example, in \cite{convolutionother}. Working with undirected graphs and the graph Laplacian rather than the adjacency matrix, \cite{convolutionother} states that ``classical modulation $\cdots$ is a multiplication by a Laplacian eigenfunction.'' But the problem with choosing as carriers the eigenvectors is that, unlike with DSP where the Hadamard product of two eigenvectors leads to another eigenvector, this is not the case in GSP. So, the product of eigenvectors does not in general lead in GSP to another eigenvector, as observed in \cite{convolutionother}, and fails to \textit{translate} the graph spectrum as desired in amplitude modulation and frequency division multiplexing. Since the eigenvectors of the shift~$\mathbf{A}$ (or of the graph Laplacian for that matter) fail the translation property, we need to look elsewhere.

Instead, we apply S$^2$CM$^2$ to modulation to develop carriers whose pointwise multiplication leads to another carrier and can be used to translate the appropriate spectral representation and be used in modulation.

\subsection{Step 1: Multiplexing in the Vertex Companion Domain}
Following DSP, we work with a companion model that preserves the cyclic nature of the DSP time model. We choose to work with the \textit{spectral} companion model\footnote{For spectral domain multiplexing, use the vertex companion model.} presented in appendix~\ref{app:spectralimpulsiverep}. The carriers are now the eigenvectors of the vertex domain shift $\mathbf{A}_\text{sp,comp}$, see equation~\eqref{eqn:Acomp}, in the \textit{spectral companion model}, \textit{not} the eigenvectors $\mathbf{v}_n$ of $\mathbf{A}$. These eigenvectors are the (conjugate) eigenfrequency vectors $\boldsymbol{\lambda}^{*^n}$, see result \ref{res:Acomp}.

The spectral companion shift $\mathbf{C}_{\textrm{sp,comp}}$ is given in equation~\eqref{eqn:compsp}. Choosing as carriers the eigenfrequency vectors $\boldsymbol{\lambda}^{*^n}$,  result~\ref{res:spshiftMC} shows that  multiplication of the signal~$\mathbf{s}$   by the GSP carriers $\lambda^{*^n}$ does shift the spectral companion representation~$\mathbf{q}_{\textrm{sp,comp}}$ by $\mathbf{C}_{\textrm{sp,comp}} = \left(\sqrt{N}\mathcal{V}^{*^{-1}}\right) \boldsymbol{\Lambda}^* \left(\frac{1}{\sqrt{N}}\mathcal{V}^*\right)$ in the spectral companion domain.

\textbf{Frequency multiplexing}. Similar to DSP frequency domain multiplexing, we then use the powers of the spectral frequency vector $\boldsymbol{\lambda}^{*^n}$ as carriers to shift and multiplex GSP signals. If the GSP signals are bandlimited in the spectral companion model, their shifted versions do not overlap and can therefore
be transmitted in a common channel.

Let $\mathbf{s}^{(0)}$, $\mathbf{s}^{(1)}$, $\cdots$, $\mathbf{s}^{(i)}$, $\cdots$, $\mathbf{s}^{(K-1)}$, be~$K$ signals that are bandlimited in the $\left\{\mathbf{q}_{\textrm{sp,comp}}\right\}$ \textit{spectral} impulse representation with bandwidth~$W$ such that $KW \leq N$. In other words, and by using~\eqref{eqn:vertexpoly}, every signal being bandlimited
\begin{align}\label{eqn:bandlimitedsi-1}
    \mathbf{s}^{(i)} {}&\xrightarrow{\left(\mathcal{V}^*\right)^{-1}} \mathbf{q}^{(i)}_{\textrm{sp,comp}} = \left[\mathbf{q}^{{\left(i_W\right)}^T}_{\textrm{sp,comp}}, 0^T\right]^T.
    \end{align}
    Further, assume all the eigenvalues of~$\mathbf{A}$, $\boldsymbol{\lambda}_k$, are non-zero.

\begin{remark}
In DSP, being bandlimited in the $\left\{\mathbf{q}_{\textrm{sp,comp}}\right\}$ spectral impulse representation is equivalent to being bandlimited in the frequency domain. In GSP, by \eqref{eqn:vertexpoly}, being bandlimited with bandlimit $W$ in the $\left\{\mathbf{q}_{\textrm{sp,comp}}\right\}$ spectral companion representation means that $\mathbf{s}$ can be written as a linear combination of $W$ lower shifted spectral functions, $\widehat{\boldsymbol{\delta}}_{sp,n}=\boldsymbol{\lambda}^{*{^n}}, n<W$.
\end{remark}

Consider a signal $\mathbf{s}$ and let the carrier signal be the spectral frequency vector power, $\boldsymbol{\lambda}^{*^i}$.

In the spectral companion model, as noted above, by result~\ref{res:spshiftMC}, multiplying~$\mathbf{s}$ by $\boldsymbol{\lambda}^{*^i}$ is equivalent to multiplying~$\mathbf{q}_{\textrm{sp,comp}}$ by $\mathbf{C}_{\textrm{sp,comp}}^i$.
\begin{equation} \label{eqn:mishift-2}
 \boldsymbol{\lambda}^{*^i} \odot \mathbf{s} \xrightarrow{\text{GFT}_{\text{comp,sp}}} \mathbf{C}_{\textrm{sp,comp}}^i \mathbf{q}_{\textrm{sp,comp}}
\end{equation}

By result~\ref{res:companionsaresame}, the spectral companion shift  $\mathbf{C}_{\textrm{sp,comp}}$ equals the vertex companion shift $\mathbf{C}_{\textrm{comp}}$. So its first lower diagonal is of ones, i.e., it contains the directed path $\mathbf{A}_{\textrm{path}}$, see equation~\eqref{eqn:structureofCcomp-2}. This shifts the signal representation~$\mathbf{q}_{\textrm{sp,comp}}$ in one-direction, exactly how DSP signals are shifted in frequency when modulated with a carrier signal, i.e., a bandlimited~$\mathbf{q}_{\textrm{sp,comp}}$ is not distorted by shifting by $\mathbf{C}_{\textrm{sp,comp}}^i$ (as long as it does not shift past the end of the path). Thus, the choice of the conjugate powers of the  eigenfrequency vector as carriers perfectly replicates the DSP modulation shift.

We modulate each signal, $\mathbf{s}^{(i)}$, using a different carrier signal, $\boldsymbol{\lambda}^{*^{W\left(i-1\right)}}$, and sum them together to produce the multiplexed signal, $\mathbf{d}_{\textrm{multipl}}$. By~\eqref{eqn:mishift-2},
\begin{align}
\label{eqn:dmult}
\mathbf{d}_{\textrm{multipl}} &= \sum_{i = 0}^{K-1} \boldsymbol{\lambda}^{*^{Wi}} \circ \mathbf{s}^{(i)} \xrightarrow{\text{GFT}_\text{comp}} \\
\mathbf{q}_{\textrm{sp,comp}}^{\mathbf{d}_{\textrm{multipl}}} &= \sum_{i = 0}^{K-1} \mathbf{C}_{\textrm{sp,comp}}^{Wi} \mathbf{q}_{\textrm{sp,comp}}^{(i)} = \left[\mathbf{q}_{\textrm{sp,comp}}^{{\left(0_B\right)}^T},\cdots, \mathbf{q}_{\textrm{sp,comp}}^{{\left({K-1}\right)_B}^T}\right]^T
\end{align}
$\mathbf{q}_{\textrm{sp,comp}}^{\mathbf{d}_{\textrm{multipl}}}$ contains the undistorted band of each signal $\mathbf{q}_{\textrm{sp,comp}}^{(i)}$ similar to the multiplexed signal in DSP.

\begin{remark}
Similar to DSP, we cannot shift $\mathbf{q}_{\textrm{sp,comp}}^{(i)}$ more than $C_{\textrm{comp}}^{W\left(K-1\right)}$. In both DSP and GSP, this would cause the band to exceed~$N$ and invoke the boundary condition/Cayley-Hamilton theorem. This would cause aliasing where the bands would overlap and sum, making $\mathbf{s}^{(i)}$ unrecoverable.
\end{remark}

To recover $\mathbf{s}^{(i)}$ from $\mathbf{q}_{\textrm{sp,comp}}^{\mathbf{d}_{\textrm{multipl}}}$, we band-pass filter the signal to extract $\mathbf{q}_{\textrm{sp,comp}}^{{\left(i_W\right)}^T}$. Then, we can demodulate the signal by $\boldsymbol{\lambda}^{*^{-Wi}}$ to shift the band back to its original position, producing the original $\mathbf{q}_{\textrm{sp,comp}}^{(i)}$. Then, using $\mathcal{V^*}$, see equations~\eqref{eqn:vertexpoly} and~\eqref{eqn:bandlimitedsi-1}, we recover the signal $\mathbf{s}^{(i)}$.

\subsection{Step 2: GSP Vertex Domain Modulation}
We convert the spectral companion model operations to the vertex GSP model.

By result \ref{res:spshiftMC}, shifting by $\mathbf{C}_{\textrm{sp,comp}}^i$ is shifting by $\textrm{M}^i$ in the spectral domain.
So, in the spectral domain, \eqref{eqn:dmult} becomes
\begin{equation}
\label{eqn:spmult}
\mathbf{d}_{\textrm{multipl}} = \sum_{i = 0}^{K-1} \boldsymbol{\lambda}^{*^{Wi}} \circ \mathbf{s}^{(i)} \xrightarrow{\text{GFT}=\mathbf{V}^{-1}} \\
\widehat{\mathbf{d}}_{\textrm{multipl}}= \sum_{i = 0}^{K-1} \mathbf{M}^{Wi} \widehat{\mathbf{s}^{(i)}}
\end{equation}
The left-hand-side of~\eqref{eqn:spmult} shows the modulated signal obtained by Hadamard product of $\mathbf{s}^{(i)}$ by the carriers $\boldsymbol{\lambda}^{*^{Wi}}$. The right-hand-side shows that modulation with these carriers $\boldsymbol{\lambda}^{*^i}$ shifts in the spectral domain the GFT's $\widehat{\mathbf{s}^{(i)}}$ of the bandlimited signals $\mathbf{s}^{(i)}$\textemdash but it is a shift using the spectral shift~$\mathbf{M}$. In general, this shifts the signal values along the edges of the spectral graph defined by~$\mathbf{M}$. Since this graph is not a directed path, this contrasts with modulation of time signals, where signals are shifted along one-direction.\footnote{\label{ftn:cyclegraphdirection} By ``one direction,'' it is meant along the direction defined by the directed edges of the cycle graph. With~$\mathbf{M}$, there is in general no ``preferred'' direction.} On the other hand, the companion model, replicating DSP and boundary conditions, produces perfect replicas (if $KW<N$) that, as just observed, the spectral domain shifting does not.

\begin{figure}[hbt!]
\begin{center}
\includegraphics[width=8.5cm, keepaspectratio]{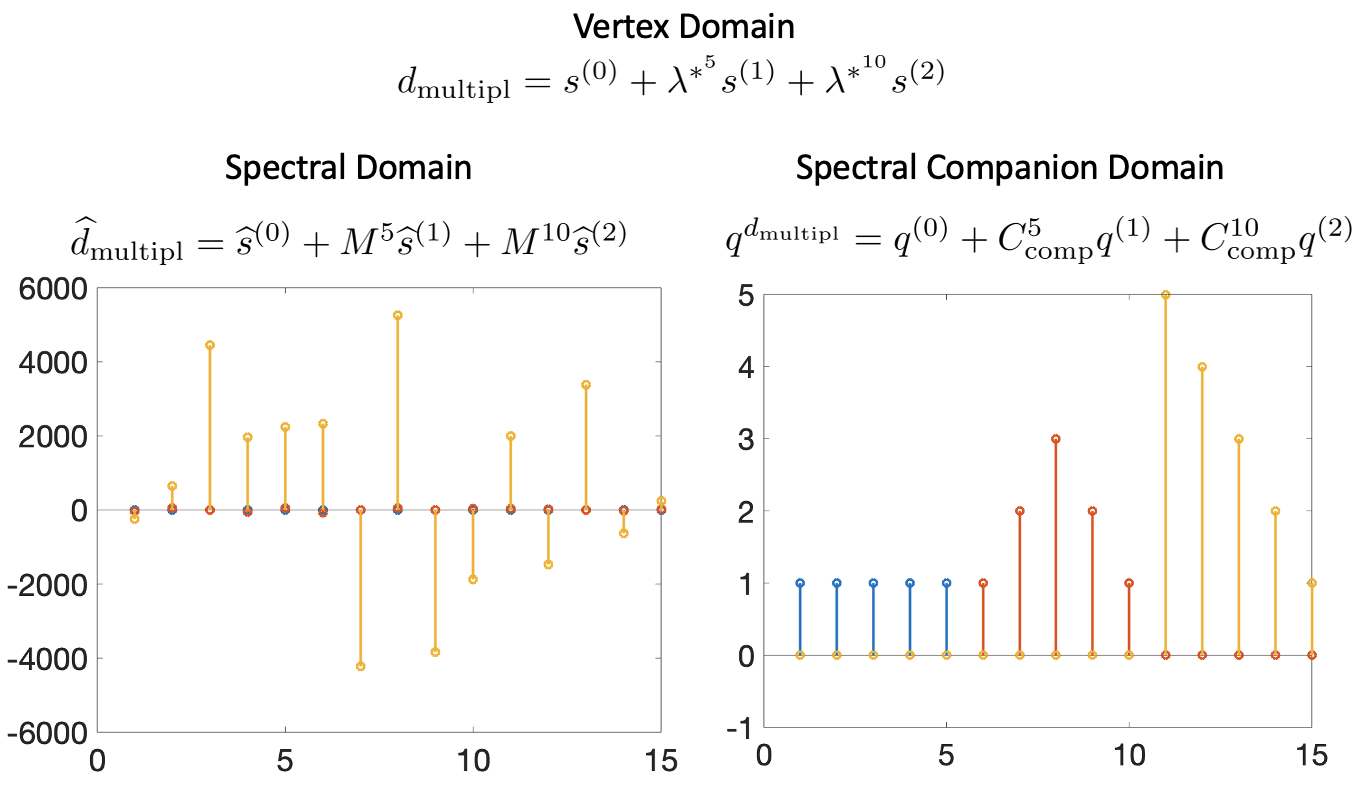}
	\caption{Modulation using \eqref{eqn:dmult}. The signal is bandlimited in the spectral and $\left\{\mathbf{q}_{\textrm{sp,comp}}^{(i)}\right\}$ \textit{spectral} companion domains, respectively. In both cases, bandwidth 5 and the bandlimited signals have following shapes:  $\mathbf{s}^{(0)}$ is rectangular, $\mathbf{s}^{(1)}$ is  triangular, and $\mathbf{s}^{(2)}$ is a ramp. The plots show the modulated signal $\mathbf{d}_{\textrm{multipl}}$ in the respective domains. For the spectral domain, see left plot, $\mathbf{d}_{\textrm{multipl}}$ is not a superposition of shifted non overlapping versions of the signals. For the  \textit{spectral} companion domain, right plot, $\left\{\mathbf{q}_{\textrm{sp,comp}}^{(i)}\right\}$ shows  perfect, shifted replicas that can be recovered.}
		\label{fig:modulation}
\end{center}
\end{figure}

We illustrate with an example. Again, we choose an undirected graph to emphasize that our results apply to both, directed and undirected graphs, working with the adjacency matrix or the graph Laplacian. We take as graph~$G$ the graph in figure~\ref{fig:mutag}, which is the undirected MUTAG graph~169, see appendix~\ref{sec:interp} for some comments on the Mutag dataset \cite{mutag}. Modulation of signals supported on graph~169 is illustrated with an example in figure~\ref{fig:modulation}. In our modulation study, we use only the graph for illustrative purposes and to exemplify modulation. We ignore the graph signal and, instead, consider for modulations three arbitrary signals indexed by the graph.
\begin{figure}[hbt!]
\begin{center}
\includegraphics[width=4cm, keepaspectratio]{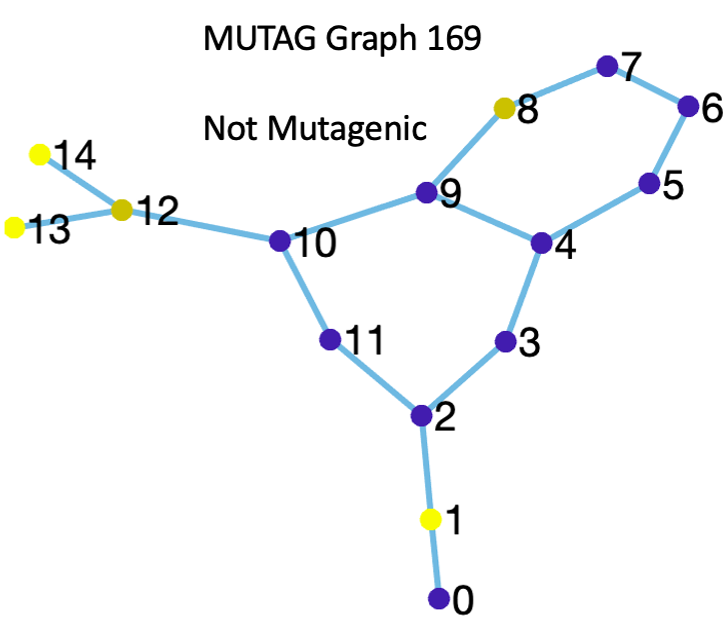}
	\caption{Graph~169 from the Mutag dataset, used only for illustration purposes.}
 \label{fig:mutag}
\end{center}
\end{figure}

On the graph~$G$ in figure~\ref{fig:mutag}, we consider three signals: $\mathbf{s}^{(0)}, \mathbf{s}^{(1)}, \mathbf{s}^{(2)}$. First, we let the signals be bandlimited in the spectral domain with bandwidth~5, where in the spectral domain   $\widehat{\mathbf{s}}^{(0)}$ is a rectangular signal, $\widehat{\mathbf{s}}^{(1)}$ is a triangular signal, and $\widehat{\mathbf{s}}^{(2)}$ is a ramp signal. We modulate the signals, forming $\mathbf{d}_{\textrm{multipl}}$ in~\eqref{eqn:dmult}. The result is shown in the spectral domain by the left plot in figure~\ref{fig:modulation}).The modulated signal, $\widehat{\mathbf{d}}_{\textrm{multipl}}$, is \textit{not} a superposition of shifted non overlapping versions of the three signals, but rather a distorted superposition of the shifted versions of the three signals.

 We repeat the process. This time the signals are bandlimited, with bandwidth~5 and the same shape as before, but in the $\left\{\mathbf{q}_{\textrm{sp,comp}}^{(i)}\right\}$ \textit{spectral} companion representation. The result is on the right side of figure~\ref{fig:modulation}, where, after modulation, $\mathbf{q}_{\textrm{sp,comp}}^{d_{\textrm{multipl}}}$ does show non distorted  shifted versions of the three original signals.

\section{conclusion}\label{sec:conclusion}
This paper introduces the \textit{companion} GSP model that explores a representation of the graph signal space as a \textit{cyclic} space, i.e., generated by a single vector and its shifts. The paper shows that the companion GSP model follows closely the DSP time model, except that it assume different boundary conditions (bc). This is captured allegorically by stating GSP=DSP+bc. Based on this, we propose the Step~1-Step~2 Companion Model Method (S$^2$CM$^2$) to expand GSP with novel concepts. In Step~1, introducing the new concept by borrowing it from DSP and taking care of the appropriate GSP model boundary conditions, and in Step~2 transport the concept to the traditional vertex and spectral GSP models through a transformation that we refer to as confusion factor. The paper illustrates S$^2$CM$^2$ with several important examples including the GSP $z$-transform, showing that convolution in the GSP companion model can be achieved with the FFT and is order $O(N\log N)$, and developing modulation and frequency multiplexing by showing that the appropriate modulating  carriers are the powers of the eigenfrequency vector (the vector that collects all the graph eigenfrequencies) and not the eigenvectors of the (vertex) shift.

Even though the paper is of a theoretical nature and the companion GSP model adds to understanding the structure of GSP models and how to expand GSP with new concepts, we describe in appendix~\ref{sec:interp} a numerical procedure to compute $\mathbf{p}_{\textrm{comp}}$ from~$\widehat{\mathbf{s}}$ by Lagrange barycentric interpolation \cite{barycentric}. The method is considered state of the art for Lagrange interpolation and works well for reasonably sized graphs. 

The paper makes throughout the assumption that the graph eigenfrequencies are distinct. This is to avoid unnecessary complications in the presentation and focus on the essential ideas. Work to be published elsewhere shows how to extend the approach to the more generic case of repeated graph eigenvalues and non diagonalizable shifts.

\bibliographystyle{ieeetr}
\bibliography{refs,sampling}
\appendices

\section{Determining the~$\mathbf{p}_{\textrm{comp}}$}\label{sec:interp}
Given the graph signal~$\mathbf{s}$, we use $\frac{1}{\sqrt{N}}\mathcal{V}\mathbf{p}_{\textrm{comp}} = \widehat{s}$ given by~\eqref{eqn:vandermonde-6} to determine the vector~$\mathbf{p}_{\textrm{comp}}$.\footnote{This method also works to determine $\mathbf{q}_{\textrm{sp,comp}}$.} Direct inversion of the Vandermonde matrix $\mathcal{V}$ is, in general, numerically unstable. On the other hand, the polynomial with vector of coefficients~$\mathbf{p}_{\textrm{comp}}$ is the polynomial interpolator across the~$N$ points $\left\{\left(\lambda_i,\sqrt{N} \widehat{s}_i\right)\right\}$. The graph eigenfrequencies $\lambda_i$ are the nodes of the interpolation. The method of choice in most cases is Lagrange polynomial interpolation implemented using barycentric formulas \cite{barycentric}.

\begin{figure}[hbt!]
\begin{center}
\includegraphics[width=8.5cm, keepaspectratio]{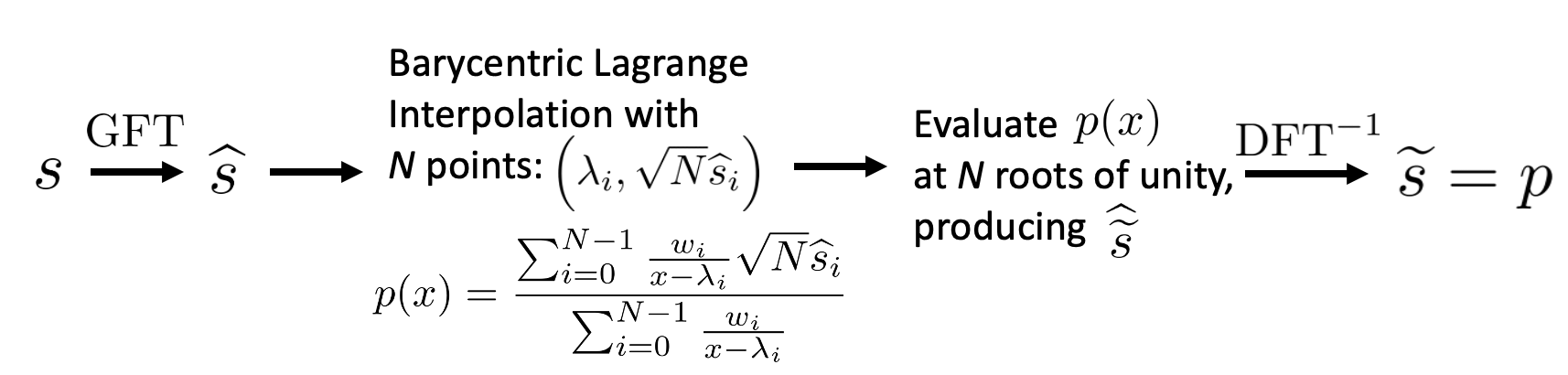}
	\caption{Method for finding $\mathbf{p}_{\textrm{comp}}$ given $\mathbf{s}$ using barycentric Lagrange interpolation with~$N$ points: $\left(\lambda_i,\sqrt{N}\widehat{s}_i\right)$. Instead of expanding the polynomial to find $\mathbf{p}_{\textrm{comp}}$, we evaluate $p(x)$ at the $N$ roots of unity to produce $\widehat{\widetilde{\mathbf{s}}}$. Using these points, we then use the $\text{DFT}^{-1}$ to find $\widetilde{\mathbf{s}} = \mathbf{p}_{\textrm{comp}}$.}
\label{fig:lagrange}
\end{center}
\end{figure}

\subsection{Barycentric Lagrange Interpolation}
Given $N$ points $\left(\lambda_i,\sqrt{N}\widehat{s}_i\right)$, let $w_i = \frac{1}{\Pi_{k\neq i}{\left(\lambda_i - \lambda_k\right)}}$. The interpolating polynomial is \cite{barycentric}:
\begin{align} \label{eqn:bary}
p(x){}& = \frac{\sum_{i=0}^{N-1}\frac{w_i}{x-\lambda_i}\sqrt{N}\widehat{s}_i}{\sum_{i=0}^{N-1}\frac{w_i}{x-\lambda_i}}
\end{align}

There are several advantages to barycentric Lagrange interpolation \cite{barycentric}: \begin{inparaenum}[1)] \item it is fast and stable; \item the numerator and denominator both have $w_i$, meaning that any common factors in $w_i$ can be cancelled, or rescaling can be used to avoid under and overflows; \item the order of the points does not matter; \item it requires $O(N^2)$ flops to calculate the weights~$\left\{w_i\right\}$ but these need to be computed only once since they do not depend on the graph signal~$\widehat{s}$; \item it requires only $O(N)$ flops to evaluate a point of $p(x)$; and \item it is forward numerically stable \cite{higham-2004}.\footnote{\label{ftn:lagrangebarystability} Another alternative form for the Lagrange interpolator is the ``modified'' Lagrange formula, e.g., \cite{barycentric,higham-2004}, that is backward stable, while the barycentric formula, and adapting the discussion in \cite{higham-2004}, ``is not backward stable, $\cdots$ but forward stable for any set of interpolating points with a small Lebesgue constant $\cdots$'' This and other advantages of the barycentric formula lead \cite{barycentric} to pick it as a `first choice.' We will not pursue further this discussion here.}
\end{inparaenum}

\subsection{Method}
To find the vector of coefficients~$\mathbf{p}_{\textrm{comp}}$ from $p(x)$, we could multiply and combine like terms in order to get $p(x)$ as in~\eqref{eqn:polyrep-1}. But this $O\left(N^2\right)$ operation is potentially numerically unstable. To find~$\mathbf{p}_{\textrm{comp}}$, we  apply a two step procedure. The first step evaluates the interpolating polynomial $p(x)$ with barycentric Lagrange interpolation~\eqref{eqn:bary} at the~$N$ roots of unity. This produces a signal $\widehat{\widetilde{\mathbf{s}}}$. Each evaluation is $O(N)$ and  numerically stable and evaluating $p(x)$ at the~$N$ roots of unity is $O\left(N^2\right)$. The second step finds the coefficients~$\mathbf{p}_{\textrm{comp}}$ of $p(x)$ by solving now a second interpolation problem, namely, of finding the interpolator $p(x)$ over the~$N$ pairs $\left(e^{-j\frac{2\pi k}{N}},\widehat{\widetilde{s}}_k\right)$. But, because the nodes of the interpolator are the $N$th-roots of unity, the associated Vandermonde is now the $\textrm{DFT}$, allowing the FFT to be used to find $\widetilde{\mathbf{s}} = \mathbf{p}_{\textrm{comp}}$. This process is illustrated in figure~\ref{fig:lagrange}.

\subsection{Example}
\begin{figure}[hbt!]
\begin{center}
	\includegraphics[width=7cm, keepaspectratio]{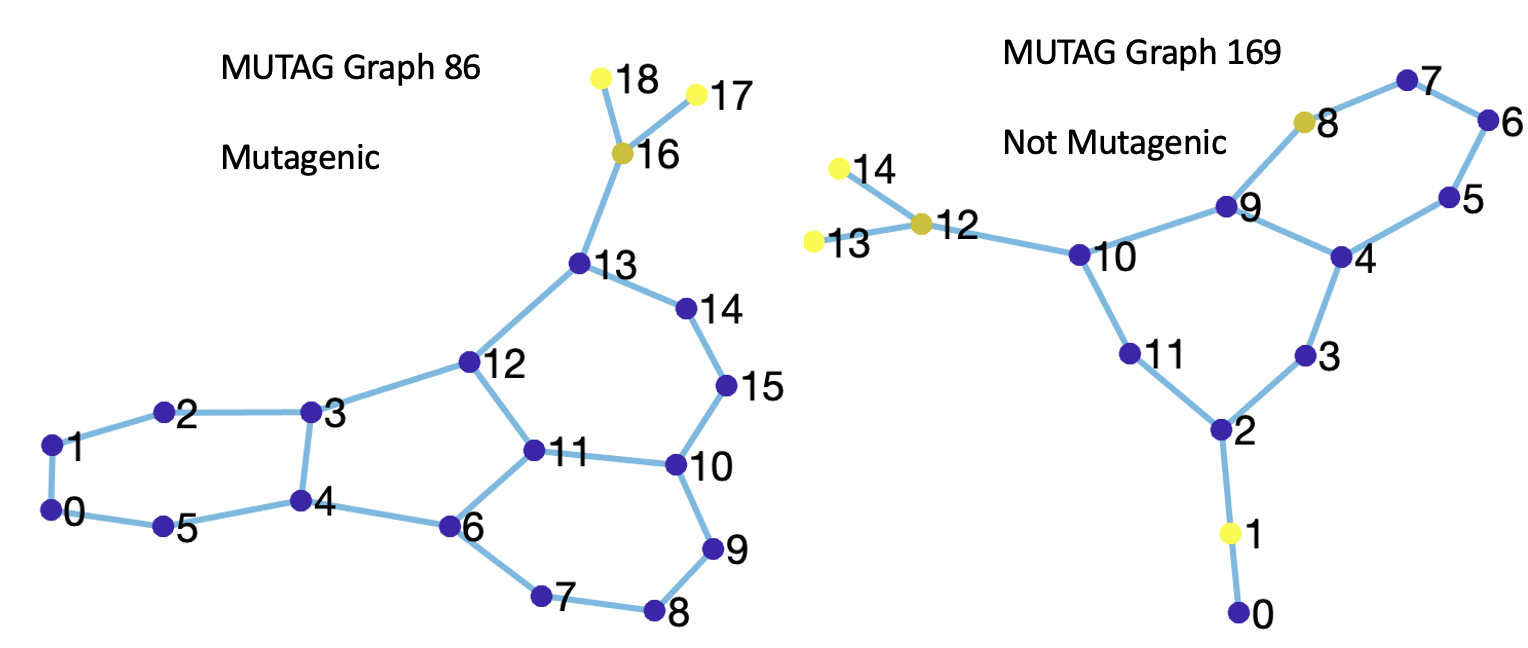}
	\caption{Graphs for two molecules in the MUTAG dataset. Graph~86 is mutagenic and graph~169 is not mutagenic. The colored nodes represent the graph signal for each graph. For graphs~86 and~169, the signals are $\mathbf{s} = [3,3,3,3,3,3,3,3,3,3,3,3,3,3,3,3,6,7,7]^T$ and $\mathbf{s} = [3,7,3,3,3,3,3,3,6,3,3,3,6,7,7]^T$, respectively.}
 \label{fig:mutag2}
\end{center}
\end{figure}

We illustrate Lagrange interpolation and its stability with two application graphs from the MUTAG dataset, see figure~\ref{fig:mutag2} \cite{mutag}. The MUTAG dataset is a set of 188 chemicals, represented by undirected graphs and divided into two classes: mutagenic and non mutagenic. The nodes are the chemical atoms and the edges are the chemical bonds.  The graph signal represents the types of chemical atoms (Carbon, Hydrogen, etc.). We apply the barycentric method to the graph signals and graphs in figure~\ref{fig:mutag2}, graphs~86 and~169 of the Mutag dataset. Note that we do not claim any practical interest from this example. It is used just to illustrate the method of computing $\mathbf{p}_{\textrm{comp}}$.

\begin{figure}[hbt!]
\begin{center}
	\includegraphics[width=6cm, keepaspectratio]{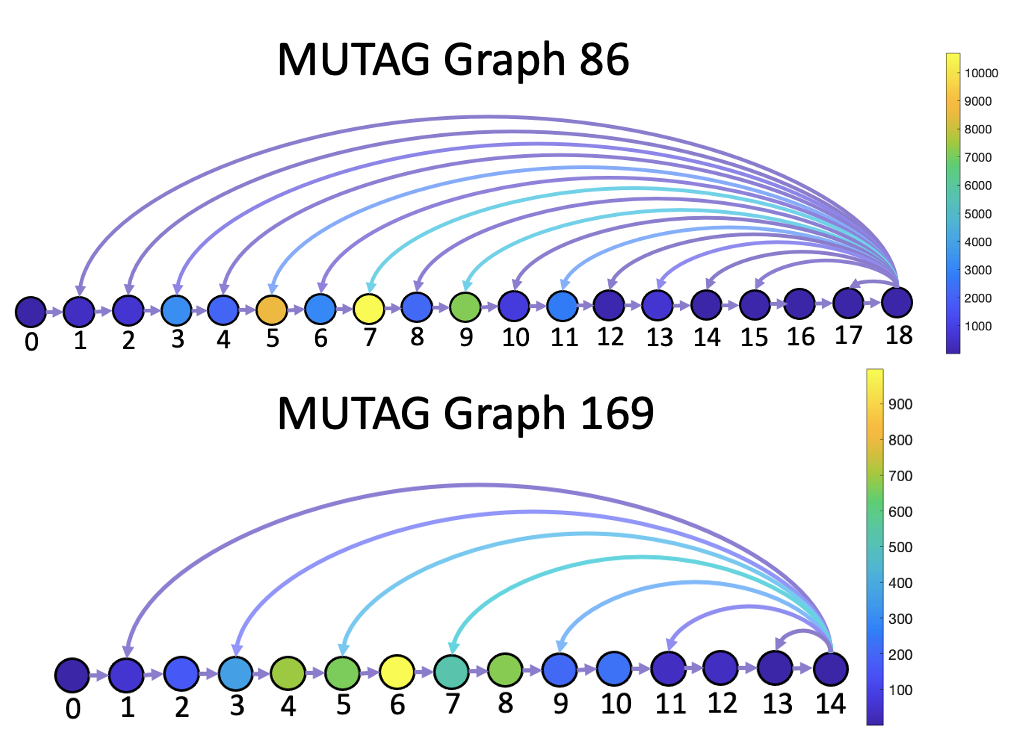}
	\caption{The signal $\mathbf{p}_{\textrm{comp}}$ (coefficients of $p(x)$) associated with $\mathbf{s}$ for two graphs in the MUTAG dataset, shown on the \textit{companion} graph for graphs~86 and~169. Note difference in color scales.}
 \label{fig:lagrangeres}
\end{center}
\end{figure}

From $\mathbf{s}$, we find $\mathbf{p}_{\textrm{comp}}$ for each graph using the method in figure~\ref{fig:lagrange}. The values of~$\mathbf{p}_{\textrm{comp}}$ are shown using a color scale on their companion graphs in figure~\ref{fig:lagrangeres}.
For Graph~86, the MSE  $\left\|\mathcal{V}\mathbf{p}_{\textrm{comp}}- \widehat{\mathbf{s}}\right\|^2\approx 0.0017$ and the MSE $\left\|\mathbf{B}_\text{comp}\boldsymbol{\delta}_0-\mathbf{s}\right\}|^2\approx 8.982\times 10^{-5}$.
For Graph~169, the MSE of $\left\|\mathcal{V}\mathbf{p}_{\textrm{comp}}- \widehat{\mathbf{s}}\right\|^2\approx 3.074\times 10^{-11}$ and the MSE $\left\|\mathbf{B}_\text{comp}\boldsymbol{\delta}_0-\mathbf{s}\right\}|^2\approx 2.049\times 10^{-11}$. This shows very good agreement and the numerically stable computation of $\mathbf{p}_{\textrm{comp}}$ for these examples.

\section{The Spectral Companion  Model}\label{app:spectralimpulsiverep}
Section~\ref{sec:gspcompanionmodel} introduced the GSP companion model starting from the vertex companion representation of section~\ref{subsubsec:companionrep} for the graph signal~$\mathbf{s}$. The vertex companion representation uses the graph impulse $\boldsymbol{\delta}_0=\textrm{GFT}^{-1}\frac{1}{\sqrt{N}}\mathbf{1}=\mathbf{V}\frac{1}{\sqrt{N}}\mathbf{1}$ and its delayed replicas $\boldsymbol{\delta}_n=\mathbf{A}^n\boldsymbol{\delta}_0$ obtained with the (vertex) graph shift~$\mathbf{A}$, see equations~\eqref{eqn:flatdelta-1} and~\eqref{eqn:shiftedelta1-6} in sections~\ref{subsubsec:graphimpulse} and~\ref{subsubsec:shiftedimpulse}. This model is needed to study modulation and frequency multiplexing in section~\ref{subsec:modulation}.

In this appendix, under assumption~\ref{ass:Adistincteigenvalues}, we develop an alternative signal companion model, the \textit{spectral companion model}, where the graph spectral signal~$\widehat{\mathbf{s}}$ plays the role of the graph signal~$\mathbf{s}$.
\subsection{The Companion Spectral Basis}\label{subsec:spcompanionbasis}
We start with the impulse.
\subsubsection{Spectral graph impulse and its shifts}\label{subsubsec:spgraphimpulse}
It uses the spectral graph impulse $\widehat{\boldsymbol{\delta}}_{{\scriptsize\textrm{sp}},0}$ and its shifts that we now introduce:
\begin{align}\label{eqn:deltasp0}
\boldsymbol{\delta}_{{\scriptsize\textrm{sp}},0}=\frac{1}{\sqrt{N}}\mathbf{1}{}&\xrightarrow{\textrm{GFT}=\mathbf{V}^{-1}}\widehat{\boldsymbol{\delta}}_{{\scriptsize\textrm{sp}},0}\:\:\Longrightarrow\:\:
   \widehat{\boldsymbol{\delta}}_{{\scriptsize\textrm{sp}},0}\delequal\mathbf{V}^{-1}\left[\frac{1}{\sqrt{N}} \mathbf{1}\right].
\end{align}
The shifts of $\widehat{\boldsymbol{\delta}}_{{\scriptsize\textrm{sp}},0}$ in the spectral domain are obtained with the spectral shift~$\mathbf{M}$, see equation~\eqref{eqn:shiftgspspmodel-1} in section~\ref{subsec:gspspmodel}. Replicating~\eqref{eqn:shiftedelta1-6}, get\footnote{\label{ftn:vertexspectralquantities}\label{rmk:vertexspectralquantities}
We will qualify by ``spectral'' any quantity related to the spectral shift~$\mathbf{M}$ or to $\boldsymbol{\delta}_{{\scriptsize\textrm{sp}},n}$, using the subscript `sp' as a reminder.}
\begin{align}\label{eqn:spdeltashifts-1}
\boldsymbol{\delta}_{{\scriptsize\textrm{sp}},n}{}&=\boldsymbol{\Lambda}^{*^{n}}\frac{1}{\sqrt{N}} \mathbf{1}=\frac{1}{\sqrt{N}}\boldsymbol{\lambda}^{*^{n}}\xrightarrow{\text{GFT}=\mathbf{V}^{-1}} \widehat{\boldsymbol{\delta}}_{{\scriptsize\textrm{sp}},n}=\mathbf{M}^n \widehat{\boldsymbol{\delta}}_{{\scriptsize\textrm{sp}},0}.
\end{align}
The left-hand-side of~\eqref{eqn:spdeltashifts-1} shows that $\boldsymbol{\delta}_{{\scriptsize\textrm{sp}},n}$ are phase shifted versions of $\frac{1}{\sqrt{N}}\mathbf{1}$ and the right-hand-side shows that their graph Fourier spectra $\widehat{\boldsymbol{\delta}}_{{\scriptsize\textrm{sp}},n}$ are shifted versions of  $\widehat{\boldsymbol{\delta}}_{{\scriptsize\textrm{sp}},0}$, but shifted by~$\mathbf{M}$.

\subsubsection{The companion spectral basis}\label{subsubsec:spectralgraphimpulse}
%
We choose as basis for the signal space module~$\mathcal{M}$
 \begin{align}
\label{eqn:graphspimpulsebasis-2}
\widehat{B}_{\scriptsize\textrm{sp,comp}}{}&\!=\!\!\left\{\widehat{\boldsymbol{\delta}}_{{\scriptsize\textrm{sp}},0},\! \cdots\!, \widehat{\boldsymbol{\delta}}_{{\scriptsize\textrm{sp}},{N-1}}\right\}
\!\!=\!\!\left\{\widehat{\boldsymbol{\delta}}_{{\scriptsize\textrm{sp}},0}, \mathbf{M}\widehat{\boldsymbol{\delta}}_{{\scriptsize\textrm{sp}},0},\!\cdots\!, \mathbf{M}^{N-1}\widehat{\boldsymbol{\delta}}_{{\scriptsize\textrm{sp}},0}\right\}\!\!.
\end{align}
Let $\widehat{\mathbf{B}}_{\scriptsize\textrm{sp,comp}}$ be the matrix whose columns are the basis vectors
\begin{align}\label{eqn:Dspimp-1}
\widehat{\mathbf{B}}_{\scriptsize\textrm{sp,comp}}{}&=\left[\begin{array}{cccc}
\widehat{\boldsymbol{\delta}}_{{\scriptsize\textrm{sp}},0}& \widehat{\boldsymbol{\delta}}_{{\scriptsize\textrm{sp}},1}&\cdots& \widehat{\boldsymbol{\delta}}_{{\scriptsize\textrm{sp}},{N-1}}
\end{array}\right].
\end{align}
Multiply each column of $\widehat{\mathbf{B}}_{\scriptsize\textrm{sp,comp}}$ by $\textrm{GFT}^{-1}=\mathbf{V}$ to get:
\begin{align}\label{eqn:graphfrequencyrepresentation-1}
\widehat{\mathbf{B}}_{\scriptsize\textrm{sp,comp}}{} \xrightarrow{\textrm{GFT}^{-1}=\mathbf{V}}\frac{1}{\sqrt{N}}\mathcal{V}^*
=\frac{1}{\sqrt{N}}\left[\!\!\!\begin{array}{cccc}1& \boldsymbol{\lambda}^{*}&\cdots& \boldsymbol{\lambda}^{*{^{N-1}}}\end{array}\!\!\!\right]
\end{align}
In~\eqref{eqn:graphfrequencyrepresentation-1}, $\boldsymbol{\lambda}$ is the eigenfrequency vector and $\mathcal{V}^*$ is the conjugate of the Vandermonde matrix that, under assumption~\ref{ass:Adistincteigenvalues}, is invertible. Hence, $\widehat{\mathbf{B}}_{\scriptsize\textrm{sp,comp}}$ is full rank, $\widehat{B}_{\scriptsize\textrm{sp,comp}}$ is a basis, and  $\widehat{\boldsymbol{\delta}}_{{\scriptsize\textrm{sp}},0}$ is a cyclic generator of the signal module~$\mathcal{M}$.
%
\subsubsection{The Companion Spectral representation}
\label{subsubsec:spectrumgraphspfreqvecrep} The spectral companion representation of $\widehat{s}$ is the coordinate vector $\mathbf{q}_{\scriptsize\textrm{sp,comp}}$ with respect to basis $\widehat{B}_{\scriptsize\textrm{sp,imp}}$:
\begin{align}
\label{eqn:shatspimpulsiverep-a1}
    \widehat{\mathbf{s}}{}&=q_0\widehat{\boldsymbol{\delta}}_{{\scriptsize\textrm{sp}},0}+q_1\widehat{\boldsymbol{\delta}}_{{\scriptsize\textrm{sp}},1}+\cdots+q_{N-1}\widehat{\boldsymbol{\delta}}_{{\scriptsize\textrm{sp}},{N-1}}\\
    \label{eqn:shatspimpulsiverep-a2}
    &=\underbrace{\left[\begin{array}{cccc}
\widehat{\boldsymbol{\delta}}_{{\scriptsize\textrm{sp}},0}&\widehat{\boldsymbol{\delta}}_{{\scriptsize\textrm{sp}},1}&\cdots&\widehat{\boldsymbol{\delta}}_{{\scriptsize\textrm{sp}},{N-1}}
\end{array}\right]}_{\widehat{\mathbf{B}}_{\scriptsize\textrm{sp,comp}}}\underbrace{\left[\begin{array}{c}
    q_0\\
    \cdots\\
    q_{N-1}
\end{array}\right]}_{\mathbf{q}_{\scriptsize\textrm{sp,imp}}}
\end{align}
Take the $\textrm{GFT}^{-1}=\mathbf{V}$ in~\eqref{eqn:shatspimpulsiverep-a1} and by equation~\eqref{eqn:graphfrequencyrepresentation-1} get~$\mathbf{s}$ in terms of $\mathbf{q}_{\textrm{sp,comp}}$.
\begin{align}\label{eqn:vertexpoly} \mathbf{s}{}&=\frac{1}{\sqrt{N}}\mathcal{V}^*\cdot \mathbf{q}_{\textrm{sp,comp}}.
\end{align}

We can get the interpretation that $\widehat{\mathbf{s}}$ is the impulse response of an LSI filter $Q_{\widehat{\mathbf{s}}}(\mathbf{M})$\textemdash a polynomial filter, not on the (vertex) shift~$\mathbf{A}$, but on the spectral shift~$\mathbf{M}$ introduced in equation~\eqref{eqn:shiftgspspmodel-1} in section~\ref{subsec:gspspmodel}.
\begin{result}[$\widehat{\mathbf{s}}$ as impulse response of $Q_{\widehat{\mathbf{s}}}(\mathbf{M})$] \label{res:widehatsimppolynomialfilterQs(M)-1}
Under assumption~\ref{ass:Adistincteigenvalues}, $\widehat{\mathbf{s}}$ is the impulse response of spectral LSI filter $Q_{\widehat{\mathbf{s}}}(\mathbf{M})$
\begin{align}\label{eqn:Q(M)companionshift-1}
    \widehat{\mathbf{s}}{}&=Q_{\widehat{\mathbf{s}}}(\mathbf{M})\widehat{\boldsymbol{\delta}}_{{\scriptsize\textrm{sp}},0}\\
\label{eqn:shatasimprespQ(M)-1}
 Q_{\widehat{\mathbf{s}}}(\mathbf{M}){}&=q_0\mathbf{I}+q_1\mathbf{M}+\cdots+q_{N-1}\mathbf{M}^{N-1}\\
\label{eqn:QsMandhats-0}
{}&=\textrm{GFT}\,\textrm{diag}\!\left[\sqrt{N}\mathbf{s}\right]\!\textrm{GFT}^{-1}\!=\!\mathbf{V}^{-1}\textrm{diag}\!\left[\sqrt{N}\mathbf{s}\right]\!\mathbf{V},
\end{align}
iff $\mathbf{q}_{\scriptsize\textup{coef}}{}=\left[\begin{array}{cccc}
q_0&
q_1&
\cdots&
q_{N-1}
\end{array}\right]^T=\mathbf{q}_{\textrm{sp,comp}}$.
\end{result}
Equation~\eqref{eqn:shatasimprespQ(M)-1} follows by successively shifting $\widehat{\boldsymbol{\delta}}_{{\scriptsize\textrm{sp}},0}$ by~$\mathbf{M}$, while~\eqref{eqn:QsMandhats-0} by factoring out the $\textrm{GFT}=\mathbf{V}^{-1}$ and its inverse in~$\mathbf{M}$ and using equation~\eqref{eqn:vertexpoly}.

\textbf{$\mathbf{q}_{\textrm{sp,comp}}$ and $\mathbf{p}_{\textrm{comp}}$}. Use~\eqref{eqn:vandermonde-5} in~\eqref{eqn:vertexpoly} and~\eqref{eqn:vandermonde-6} in~\eqref{eqn:shatspimpulsiverep-a2} to get
\begin{alignat}{2}\label{eqn:pimpandqspimp}
\mathbf{q}_{\textrm{sp,comp}}{}&=\left(\mathcal{V}^*\right)^{-1}\textrm{GFT}^{-1}\mathcal{V} \, \mathbf{p}_{\textrm{comp}}{}&=\left(\mathcal{V}^*\right)^{-1}\mathbf{V}\mathcal{V}\mathbf{p}_{\textrm{comp}}\\
\label{eqn:qspimpandpimp}
\mathbf{p}_{\textrm{comp}}{}&=\mathcal{V}^{-1}\,\,\textrm{GFT}\,\,\mathcal{V}^*\,\mathbf{q}_{\textrm{sp,comp}}\,\,{}&=\mathcal{V}^{-1}\mathbf{V}^{-1}\,\mathcal{V}^*\mathbf{q}_{\textrm{sp,comp}}.
\end{alignat}

\subsection{Spectral Companion Shift and Graph}\label{subsec:spectralcompanionmodel}
Following section~\ref{subsec:companionshift} but now using $\mathbf{M}$ and $\widehat{\boldsymbol{\delta}}_{\textrm{sp,comp}}$, get 
\begin{align} \label{eqn:compsp-1}
\mathcal{M}\cdot\widehat{\mathbf{s}}{}&=\mathcal{M}\cdot\widehat{\mathbf{B}}_{\textrm{sp,comp}}\cdot\mathbf{q}_{\textrm{sp,comp}}=\widehat{\mathbf{B}}_{\textrm{sp,comp}}\cdot\mathbf{C}_{\textrm{sp,comp}}\cdot\mathbf{q}_{\textrm{sp,comp}}
\end{align}
Since true for every $\mathbf{q}_{\textrm{sp,comp}}$, and by diagonalization of~$\mathbf{M}$, get
\begin{align}
\label{eqn:compsp-2}
{}&\Rightarrow\mathbf{C}_{\textrm{sp,comp}}=\left(\mathbf{V}\cdot\widehat{\mathbf{B}}_{\textrm{sp,comp}}\right)^{-1}\boldsymbol{\Lambda}^*\mathbf{V}\cdot\widehat{\mathbf{B}}_{\textrm{sp,comp}}\\
\label{eqn:compsp}
&\xrightarrow{\textrm{by~\eqref{eqn:graphfrequencyrepresentation-1}}}\mathbf{C}_{\scriptsize\textrm{sp,comp}}= \sqrt{N}\mathcal{V}^{*^{-1}} \boldsymbol{\Lambda}^* \frac{1}{\sqrt{N}}\mathcal{V}^*=\mathcal{V}^{*^{-1}} \boldsymbol{\Lambda}^* \mathcal{V}^*
\end{align}
The spectral companion shift $\mathbf{C}_{\scriptsize\textrm{sp,comp}}$ is a companion matrix.

\begin{result}[Modulation and shifting] \label{res:spshiftMC}
\begin{inparaenum}[1)]
\item Modulating~$\mathbf{s}$ by $\boldsymbol{\lambda}^*$, i.e., $\boldsymbol{\lambda}^*\circ \mathbf{s}$, shifts~$\widehat{\mathbf{s}}$ by~$\mathbf{M}$ in the spectral domain.
\item Shifting $\widehat{\mathbf{s}}$ by $\mathbf{M}$ shifts $\mathbf{q}_{\textrm{sp,comp}}$ by $\mathbf{C}_{\scriptsize\textrm{sp,comp}}$.
\end{inparaenum}
\end{result}
The first part is in \cite{shimoura-2021} and the second part is from~\eqref{eqn:compsp-1}.
\begin{result}[Vertex domain shift $\mathbf{A}_{\textrm{sp,comp}}$ in the spectral companion model]\label{res:Acomp} The vertex domain shift $\mathbf{A}_{\textrm{sp,comp}}$ and its eigenvectors in the spectral companion model are
\begin{align}\label{eqn:Acomp}
\mathbf{A}_{\textrm{sp,comp}}{}&=\frac{1}{\sqrt{N}}\mathcal{V}^* \boldsymbol{\Lambda} \sqrt{N}\mathcal{V}^{*^{-1}}\\
\label{eqn:Acompeigenvectors}
\frac{1}{\sqrt{N}}\boldsymbol{\lambda}^{*^n}, n{}&=0,\cdots, N-1
\end{align}
\end{result}
\begin{proof}
By equation~\eqref{eqn:compsp},
 the shift in the spectral domain of the spectral companion model is $\mathbf{C}_{\textrm{sp,comp}}$. By exchanging $\frac{1}{\sqrt{N}}\mathcal{V}^*$ and $\sqrt{N}\mathcal{V}^{*^{-1}}$ in the diagonalization of $\mathbf{C}_{\textrm{sp,comp}}$ in~\eqref{eqn:compsp}, obtain $\mathbf{A}_{\textrm{sp,comp}}$ as in result~\ref{res:Acomp}. The eigenvectors of $\mathbf{A}_{\textrm{sp,comp}}$ are the columns of $\frac{1}{\sqrt{N}}\mathcal{V}^*$.
\end{proof}

The spectral companion graph $G_{\textrm{sp,comp}}$ has as adjacency matrix the spectral companion shift $\mathbf{C}_{\scriptsize\textrm{sp,comp}}$.
\subsection{GFT in the Spectral Companion Model}\label{subsec:gftinspcompanionmodel}
We next obtain the graph Fourier transform in the spectral companion model.
\begin{result}[Spectral companion graph Fourier transform $\textrm{GFT}_{\textrm{sp,comp}}$]\label{res:GFTcomp,sp}
 For the spectral companion model,
\begin{align}\label{eqn:GFTcompsp}
\textrm{GFT}_{\textrm{sp,comp}}{}&=\sqrt{N}\mathcal{V}^{*^{-1}}\\
\label{eqn:GFTcompsp-2}
\mathbf{s}&\xrightarrow{{\textrm{GFT}}_{\textrm{sp,comp}}=\sqrt{N}\mathcal{V}^{*^{-1}}} \mathbf{q}_{\textrm{sp,comp}}=\sqrt{N}\mathcal{V}^{*^{-1}} \mathbf{s}.
\end{align}
\end{result}
\begin{proof}
Obtain~\eqref{eqn:GFTcompsp} from~\eqref{eqn:compsp} and~\eqref{eqn:GFTcompsp-2} from equation~\eqref{eqn:vertexpoly}.
\end{proof}
By result~\ref{res:GFTcomp,sp}, in the spectral companion model~$\mathbf{s}$ and~$\mathbf{q}_{\textrm{sp,comp}}$ are a Fourier pair.

\begin{result}\label{res:companionsaresame}
The companion matrix and companion graph for the vertex and spectral companion models are the same.
\end{result}
\begin{proof}
Since~$\mathbf{A}$ and~$\mathbf{M}$ are co-spectral\footnote{Matrices~$\mathbf{A}$ and~$\mathbf{M}$ are real valued so their complex eigenvalues occur in conjugate pairs and $\boldsymbol{\Lambda}^*$ and $\boldsymbol{\Lambda}$ contain the same entries.},  $\Delta_\mathbf{A}(x)=\Delta_\mathbf{M}(x)$. So, $\mathbf{C}_{\scriptsize\textrm{comp}}=\mathbf{C}_{\scriptsize\textrm{sp,comp}}$ and, thus, define the same companion graph $G_{\scriptsize\textrm{comp}}$ in section~\ref{subsec:companiongraph}.
\end{proof}

\end{document}

%% file: prologue.tex
\usepackage{cite}

\usepackage{amsmath,amssymb,amsfonts}
\usepackage{mathtools}

\usepackage{stfloats}
\usepackage{array}

\usepackage{graphicx}
\usepackage{textcomp}
\usepackage{xcolor}
\usepackage{mathrsfs}
\usepackage{amsthm}
\usepackage{scalerel,stackengine}
\usepackage{paralist}
\usepackage{blkarray}
\usepackage[ruled,vlined,linesnumbered]{algorithm2e}
\usepackage{empheq}
\definecolor{lightgreen}{HTML}{90EE90}

\newcolumntype{P}[1]{>{\centering\arraybackslash}p{#1}}

\usepackage[most]{tcolorbox}

\tcbset{colback=blue!3!white, colframe=red!50!black,
        highlight math style= {enhanced, 
     colframe=red,colback=red!10!white,boxsep=0pt}
       }

    \usepackage[compact]{titlesec}
    \titlespacing{\section}{0pt}{2.5ex}{1ex}
    \titlespacing{\subsection}{0pt}{1ex}{0ex}

\usepackage{caption}
\usepackage{newtxmath}
\usepackage{alltt}
\usepackage{epsfig}
\usepackage{verbatim}
\usepackage{fancybox}
\usepackage{subfigure}
\usepackage{url}
\usepackage{multirow}
\usepackage{setspace}
\usepackage{fancyhdr}
\usepackage{framed}
\usepackage{booktabs}
\usepackage{color}
\usepackage{wrapfig}

\usepackage{mathtools}
\mathtoolsset{showonlyrefs}
\usepackage{placeins}

\usepackage{tikz}
\usepackage{enumitem}
\setitemize{noitemsep,topsep=0pt,parsep=0pt,partopsep=0pt}

\newcommand\oast{\stackMath\mathbin{\stackinset{c}{0ex}{c}{0ex}{\ast}{\bigcirc}}}
\def\delequal{\mathrel{\ensurestackMath{\stackon[1pt]{=}{\scriptstyle\Delta}}}}

\allowdisplaybreaks

\newtheorem{remark}{Remark}
\newtheorem{example}{Example}

\newtheorem{assumption}{Assumption}
\newtheorem{result}{Result}

\newcommand{\xdownarrow}[1]{%
{\frac{}{}\left\downarrow\vbox to #1{}\right.\kern-\nulldelimiterspace}
}
\newcommand{\xuparrow}[1]{%
{\left\uparrow\vbox to #1{}\right.\kern-\nulldelimiterspace}
}
\newcommand{\xupdownarrow}[1]{%
{\left\updownarrow\vbox to #1{}\right.\kern-\nulldelimiterspace}
}

\newcommand{\xydownarrow}[2][]{%
\left.{#1}\right\downarrow{#2}} 

%% file: main.bbl
\begin{thebibliography}{10}

\bibitem{Sandryhaila:13}
A.~Sandryhaila and J.~M.~F. Moura, ``Discrete signal processing on graphs,''
  {\em IEEE Trans. Signal Proc.}, vol.~61, pp.~1644--1656, January 2013.

\bibitem{Sandryhaila:14}
A.~Sandryhaila and J.~M.~F. Moura, ``Discrete signal processing on graphs:
  Frequency analysis,'' {\em IEEE Trans. Signal Proc.}, vol.~62,
  pp.~3042--3054, June 2014.

\bibitem{ShumanNFOV:13}
D.~I. Shuman, S.~K. Narang, P.~Frossard, A.~Ortega, and P.~Vandergheynst, ``The
  emerging field of signal processing on graphs: {E}xtending high-dimensional
  data analysis to networks and other irregular domains,'' {\em IEEE Signal
  Proc. Magazine}, vol.~30, pp.~83--98, May 2013.

\bibitem{oppenheimwillsky-1983}
A.~V. Oppenheim and A.~S. Willsky, {\em Signals and Systems}.
\newblock Englewood Cliffs, New Jersey: Prentice-Hall, 1983.

\bibitem{siebert-1986}
W.~M. Siebert, {\em Circuits, Signals, and Systems}.
\newblock Cambridge, MA: The MIT Press, 1986.

\bibitem{oppenheimschaffer-1989}
A.~V. Oppenheim and R.~W. Schafer, {\em Discrete-Time Signal Processing}.
\newblock Englewood Cliffs, New Jersey: Prentice-Hall, 1989.

\bibitem{mitra-1998}
S.~K. Mitra, {\em Digital Signal Processing. A Computer-Based Approach}.
\newblock New York: McGraw Hill, 1998.

\bibitem{Sandryhaila:14big}
A.~Sandryhaila and J.~M.~F. Moura, ``Big data analysis with signal processing
  on graphs: Representation and processing of massive data sets with irregular
  structure,'' {\em IEEE Signal Processing Magazine}, vol.~31, pp.~80--90,
  September 2014.

\bibitem{Pueschel:03a}
M.~P{\"u}schel and J.~M.~F. Moura, ``The algebraic approach to the discrete
  cosine and sine transforms and their fast algorithms,'' {\em SIAM J. Comp.},
  vol.~32, pp.~1280--1316, May 2003.

\bibitem{Pueschel:05e}
M.~P{\"u}schel and J.~M.~F. Moura, ``Algebraic signal processing theory.'' 67
  pages., December 2006.

\bibitem{Pueschel:08a}
M.~P{\"u}schel and J.~M.~F. Moura, ``Algebraic signal processing theory:
  Foundation and {1-D} time,'' {\em IEEE Trans. Signal Proc.}, vol.~56,
  pp.~3572--3585, August 2008.

\bibitem{Pueschel:08b}
M.~P{\"u}schel and J.~M.~F. Moura, ``Algebraic signal processing theory: {1-D}
  space,'' {\em IEEE Trans. Signal Proc.}, vol.~56, pp.~3586--3599, August
  2008.

\bibitem{Pueschel:08c}
M.~P{\"u}schel and J.~M.~F. Moura, ``Algebraic signal processing theory:
  {Cooley}-{Tukey} type algorithms for {DCT}s and {DST}s,'' {\em IEEE Trans.
  Signal Proc.}, vol.~56, pp.~1502--1521, April 2008.

\bibitem{Chung:96}
F.~R.~K. Chung, {\em Spectral Graph Theory}.
\newblock AMS, 1996.

\bibitem{Belkin:02}
M.~Belkin and P.~Niyogi, ``Using manifold structure for partially labeled
  classification,'' in {\em Neural Information Processing Symposium (NIPS)},
  2002.

\bibitem{Coifman:05a}
R.~R. Coifman, S.~Lafon, A.~Lee, M.~Maggioni, B.~Nadler, F.~J. Warner, and
  S.~W. Zucker, ``Geometric diffusions as a tool for harmonic analysis and
  structure definition of data: {D}iffusion maps,'' {\em Proc. Nat. Acad.
  Sci.}, vol.~102, no.~21, pp.~7426--7431, 2005.

\bibitem{guestrinbodikthibauxpaskinmadden-ipsn2004}
C.~Guestrin, P.~Bodik, R.~Thibaux, M.~Paskin, and S.~Madden, ``Distributed
  regression: an efficient framework for modeling sensor network data,'' in
  {\em IPSN}, pp.~1--10, 2004.

\bibitem{wagnerbaraniuketal-SSPWorkshop2005}
R.~Wagner, H.~Choi, R.~G. Baraniuk, and V.~Delouille, ``Distributed wavelet
  transform for irregular sensor network grids,'' in {\em IEEE SSP Workshop},
  pp.~1196--1201, 2005.

\bibitem{Hammond:11}
D.~K. Hammond, P.~Vandergheynst, and R.~Gribonval, ``Wavelets on graphs via
  spectral graph theory,'' {\em J. Appl. Comp. Harm. Anal.}, vol.~30, no.~2,
  pp.~129--150, 2011.

\bibitem{Narang:12}
S.~K. Narang and A.~Ortega, ``Perfect reconstruction two-channel wavelet filter
  banks for graph structured data,'' {\em IEEE Trans. Signal Proc.}, vol.~60,
  no.~6, pp.~2786--2799, 2012.

\bibitem{Narang:10}
S.~K. Narang and A.~Ortega, ``Local two-channel critically sampled filter-banks
  on graphs,'' in {\em ICIP}, pp.~333--336, 2010.

\bibitem{Narang:11}
S.~K. Narang and A.~Ortega, ``Downsampling graphs using spectral theory,'' in
  {\em IEEE International Conference on Acoustics, Speech and Signal Processing
  (ICASSP)}, pp.~4208--4211, 2011.

\bibitem{ortegafrossardkovacevicmouravandergheynst-2018}
A.~{Ortega}, P.~{Frossard}, J.~{Kovačević}, J.~M.~F. {Moura}, and
  P.~{Vandergheynst}, ``Graph signal processing: Overview, challenges, and
  applications,'' {\em Proceedings of the IEEE}, vol.~106, pp.~808--828, May
  2018.

\bibitem{leusmarquesmouraortegashuman}
G.~Leus, A.~G. Marques, J.~M. Moura, A.~Ortega, and D.~I. Shuman, ``Graph
  signal processing: History, development, impact, and outlook,'' {\em IEEE
  Signal Processing Magazine}, vol.~40, no.~4, pp.~49--60, 2023.

\bibitem{giraultgoncalvesfleury-2015}
B.~Girault, P.~Gonçalves, and {\'E}.~Fleury, ``Translation on graphs: An
  isometric shift operator,'' {\em IEEE Signal Processing Letters}, vol.~22,
  pp.~2416--2420, Dec 2015.

\bibitem{gavilizhang-2017}
A.~Gavili and X.~P. Zhang, ``On the shift operator, graph frequency, and
  optimal filtering in graph signal processing,'' {\em IEEE Transactions on
  Signal Processing}, vol.~65, pp.~6303--6318, Dec 2017.

\bibitem{Zhu:12}
X.~Zhu and M.~Rabbat, ``Approximating signals supported on graphs,'' in {\em
  IEEE International Conference on Acoustics, Speech and Signal Processing
  (ICASSP)}, pp.~3921--3924, 2012.

\bibitem{anis2014towards}
A.~Anis, A.~Gadde, and A.~Ortega, ``Towards a sampling theorem for signals on
  arbitrary graphs,'' in {\em 2014 IEEE Int. Conf. on Acoustics, Speech and
  Signal Processing (ICASSP)}, pp.~3864--3868, IEEE, 2014.

\bibitem{Jelena}
S.~Chen, R.~Varma, A.~Sandryhaila, and J.~Kova\v{c}evi\'c, ``Discrete signal
  processing on graphs: Sampling theory,'' {\em IEEE Trans. Signal Proc.},
  vol.~63, no.~24, pp.~6510 -- 6523, 2015.

\bibitem{marques2015sampling}
A.~G. Marques, S.~Segarra, G.~Leus, and A.~Ribeiro, ``Sampling of graph signals
  with successive local aggregations,'' {\em IEEE Transactions on Signal
  Processing}, vol.~64, no.~7, pp.~1832--1843, 2015.

\bibitem{anis2016efficient}
A.~Anis, A.~Gadde, and A.~Ortega, ``Efficient sampling set selection for
  bandlimited graph signals using graph spectral proxies,'' {\em IEEE Trans.
  Signal Proc.}, vol.~64, no.~14, pp.~3775--3789, 2016.

\bibitem{chamon2017greedy}
L.~F. Chamon and A.~Ribeiro, ``Greedy sampling of graph signals,'' {\em IEEE
  Transactions on Signal Processing}, vol.~66, no.~1, pp.~34--47, 2017.

\bibitem{tanaka-2018}
Y.~{Tanaka}, ``Spectral domain sampling of graph signals,'' {\em IEEE
  Transactions on Signal Processing}, vol.~66, no.~14, pp.~3752--3767, 2018.

\bibitem{shimoura-2021}
J.~Shi and J.~M.~F. Moura, ``Graph signal processing: Dualizing {GSP} sampling
  in the vertex and spectral domains,'' {\em IEEE Transactions on Signal
  Processing}, vol.~70, pp.~2883--2898, 2022.

\bibitem{EldarTanakaSPM}
Y.~Tanaka, Y.~C. Eldar, A.~Ortega, and G.~Cheung, ``Sampling signals on graphs:
  From theory to applications,'' {\em IEEE Signal Processing Magazine},
  vol.~37, no.~6, pp.~14--30, 2020.

\bibitem{tekevaidyanathan-2017-I}
O.~{Teke} and P.~P. {Vaidyanathan}, ``Extending classical multirate signal
  processing theory to graphs—part~{I} fundamentals,'' {\em IEEE Transactions
  on Signal Processing}, vol.~65, no.~2, pp.~409--422, 2017.

\bibitem{tekevaidyanathan-2017-II}
O.~{Teke} and P.~P. {Vaidyanathan}, ``Extending classical multirate signal
  processing theory to graphs—part~{II}: {M}-channel filter banks,'' {\em
  IEEE Transactions on Signal Processing}, vol.~65, no.~2, pp.~423--437, 2017.

\bibitem{Agaskar:12}
A.~Agaskar and Y.~M. Lu, ``A spectral graph uncertainty principle,'' {\em IEEE
  Transactions on Information Theory}, vol.~59, no.~7, pp.~4338--4356, 2013.

\bibitem{pasdeloupgriponmercierpastorrabbat-2018}
B.~{Pasdeloup}, V.~{Gripon}, G.~{Mercier}, D.~{Pastor}, and M.~G. {Rabbat},
  ``Characterization and inference of graph diffusion processes from
  observations of stationary signals,'' {\em IEEE Transactions on Signal and
  Information Processing over Networks}, vol.~4, no.~3, pp.~481--496, 2018.

\bibitem{chen2015signal-2}
S.~Chen, A.~Sandryhaila, J.~M.~F. Moura, and J.~Kova{\v{c}}evi{\'c}, ``Signal
  recovery on graphs: Variation minimization,'' {\em IEEE Transactions on
  Signal Processing}, vol.~63, no.~17, pp.~4609--4624, 2015.

\bibitem{segarra2015interpolation}
S.~Segarra, A.~G. Marques, G.~Leus, and A.~Ribeiro, ``Interpolation of graph
  signals using shift-invariant graph filters,'' in {\em 2015 23rd European
  Signal Processing Conference (EUSIPCO)}, pp.~210--214, IEEE, 2015.

\bibitem{segarra2016reconstruction}
S.~Segarra, A.~G. Marques, G.~Leus, and A.~Ribeiro, ``Reconstruction of graph
  signals through percolation from seeding nodes,'' {\em IEEE Transactions on
  Signal Processing}, vol.~64, no.~16, pp.~4363--4378, 2016.

\bibitem{convolutionother}
D.~I. Shuman, B.~Ricaud, and P.~Vandergheynst, ``Vertex-frequency analysis on
  graphs,'' {\em Appl. Comput. Harmon. Anal.}, vol.~40, no.~2, pp.~260--291,
  2016.

\bibitem{marquessegarraleusribeiro-17}
A.~G. {Marques}, S.~{Segarra}, G.~{Leus}, and A.~{Ribeiro}, ``Stationary graph
  processes and spectral estimation,'' {\em IEEE Transactions on Signal
  Processing}, vol.~65, no.~22, pp.~5911--5926, 2017.

\bibitem{meimoura-2017}
J.~{Mei} and J.~M.~F. {Moura}, ``Signal processing on graphs: Causal modeling
  of unstructured data,'' {\em IEEE Transactions on Signal Processing},
  vol.~65, no.~8, pp.~2077--2092, 2017.

\bibitem{meimoura-2018}
J.~{Mei} and J.~M.~F. {Moura}, ``Silvar: Single index latent variable models,''
  {\em IEEE Trans. on Signal Processing}, vol.~66, no.~11, pp.~2790--2803,
  2018.

\bibitem{dongthanourabbatfrossard-2019}
X.~{Dong}, D.~{Thanou}, M.~{Rabbat}, and P.~{Frossard}, ``Learning graphs from
  data: A signal representation perspective,'' {\em IEEE Signal Processing
  Magazine}, vol.~36, no.~3, pp.~44--63, 2019.

\bibitem{derimoura-2017}
J.~A. Deri and J.~M.~F. Moura, ``Spectral projector-based graph {F}ourier
  transforms,'' {\em IEEE Journal of Selected Topics in Signal Processing},
  vol.~11, pp.~785--795, September 2017.

\bibitem{shimoura-asilomar2019}
J.~Shi and J.~M.~F. Moura, ``Topics in graph signal processing: Convolution and
  modulation,'' in {\em 2019 53rd Asilomar Conference on Signals, Systems, and
  Computers}, pp.~457--461, 2019.

\bibitem{gantmacher1959matrix}
F.~R. Gantmacher, ``Matrix theory,'' {\em Chelsea, New York}, vol.~21, 1959.

\bibitem{hoffmankunze}
K.~Hoffman and R.~Kunze, {\em Linear algebra}.
\newblock Prentice-Hall Mathematics Series, NJ: Englewood Cliffs, Prentice
  Hall, Inc, 1961.

\bibitem{Knuth81}
D.~E. Knuth, {\em The Art of Computer Programming}, vol.~2: Seminumerical
  Algorithms.
\newblock Addison-Wesley, 1981.

\bibitem{mutag}
R.~A. Rossi and N.~K. Ahmed, ``The network data repository with interactive
  graph analytics and visualization,'' in {\em AAAI}, 2015.

\bibitem{barycentric}
J.-P. Berrut and L.~N. Trefethen, ``Barycentric lagrange interpolation,'' {\em
  SIAM Review}, vol.~46, no.~3, pp.~501--517, 2004.

\bibitem{higham-2004}
N.~J. Higham, ``The numerical stability of barycentric {L}agrange
  interpolation,'' {\em IMA Journal of Numerical Analysis}, vol.~24, no.~4,
  p.~547–556, 2004.

\end{thebibliography}
